\documentclass[11pt, a4paper]{article}
\usepackage[pdfauthor={P. P. Cook},
pdftitle={Exotic $E_{11}$ branes as composite gravitational solutions}]{hyperref}
\usepackage{amsmath}
\usepackage{graphicx}
\usepackage{youngtab}
\usepackage{amssymb}
\usepackage{amsthm}
\usepackage{chngcntr}
\usepackage{array}
\usepackage{xy}
\counterwithin{figure}{section}
\counterwithin{table}{section}
\newcommand{\comment}[1]{}
\renewcommand{\vec}[1]{{\underline{#1}}}

\begin{document}
\title{Exotic $E_{11}$ branes as composite gravitational solutions}
\author{Paul P. Cook\footnote{email: paul.cook@kcl.ac.uk} \\ 
\\
{\itshape Department of Mathematics, King's College London \\ 
The Strand, London WC2R 2LS, UK}
}
\begin{titlepage}
\begin{flushright}
KCL-MTH-09-07 \\
%{\tt hep-th/????.????}
\end{flushright}
\vspace{70pt}
\centering{\LARGE Exotic $E_{11}$ branes as composite gravitational solutions}\\
\vspace{30pt}
\def\thefootnote{\fnsymbol{footnote}}
Paul P. Cook\footnote{\href{mailto:paul.cook@kcl.ac.uk}{email: paul.cook@kcl.ac.uk}}\\
\setcounter{footnote}{0}

\vspace{10pt}
{\itshape Department of Mathematics, King's College London \\ 
The Strand, London WC2R 2LS, UK}\\
\vspace{30pt}
\begin{abstract}
A two-parameter group element is presented that interpolates between M-brane solutions. The group element is used to interpret a number of exotic branes related to the generators of the adjoint representation of $E_{11}$ as non-marginal half-BPS bound states of M-branes. It is conjectured that the adjoint representation of $E_{11}$ contains only generators related to bound states of fundamental M-branes which, in the limit, may be understood as membrane molecules.
\end{abstract}
\end{titlepage}
\clearpage
\newpage

\begin{section}{Introduction}
The fundamental solutions of M-theory, the KK-wave \cite{Brinkmann:1923p2723}, the membrane \cite{Duff:1991p2619}, the fivebrane \cite{Gueven:1992p2620} and the KK-monopole \cite{Gross:1983p687,Sorkin:1983p2622} were known as solutions of eleven dimensional supergravity or, even, general relativity, prior to the first investigations of M-theory. Solutions particular to M-theory are thin on the ground. More intricate solutions of M-theory can be constructed from the fundamental solutions either as marginal brane intersections or as non-marginal bound states. These solutions are valid beyond the weak-coupling limit of M-theory and are solutions that probe the nature of M-theory's extension of supergravity. 
 
Of these two classes of solution the M-brane intersections \cite{Papadopoulos:1996p1472} have been classified (for reviews see \cite{Smith:2002p1483,Gauntlett:1997p1516} and the references therein), as have the brane intersection rules of their counterparts with Euclidean worldvolumes, the S-branes \cite{Deger:2002p1845,Ohta:2003p1840}. The basic brane intersections are $M2\perp M2(0)$, $M2\perp M5(1)$, $M5\perp M5(3)$\footnote{This notation indicates the pair of branes involved in the intersecton, while the number in brackets indicates the number of worldvolume directions along the intersections, so that $M2 \perp M2(0)$ indicates two M2 branes intersecting at a point.} and intersections involving the KK-wave and KK-monopole are also known \cite{Tseytlin:1996p1469,Costa:1996p1651,Bergshoeff:1997p694}. A marginal brane intersection involving N different branes is typically described by N harmonic functions and preserves $\frac{1}{2^N}$ of the background supersymmetry. Although classified, the exact forms of the harmonic functions encoding localised intersecting solutions are not known in a closed form. In IIA supergravity there is an integral form of a localised solution of D2 branes localised in the worldvolume of D6 branes \cite{Cherkis:2002p1984} and this construction technique, relying on the special property that D6 branes uplift to a pure gravitational solution, remains the state of the art method.

However marginal solutions are easy to construct from two branes under less general conditions:
\begin{list}{$\bullet$}{}
\item {when both branes are smeared in their relative transverse directions \cite{Tseytlin:1996p1469}};
\item {when one of the branes is smeared (e.g. for an M2 (012) brane intersecting a second M2 (034) brane at a point, where the harmonic function encoding one of the brane solutions is a function of only the mutual transverse coordinates of both branes)}; and
\item {in the near-core limit, where an exact solution to the harmonic functions can be computed in the near horizon limit (or equivalently, a large charge limit) of one of the branes.}
\end{list}
In contrast the non-marginal, bound states of M-theory have not been catalogued, but a number of such solutions have been found. The first discovered of these was named the dyonic membrane \cite{Izquierdo:1995p1636} and was lifted from N=2 D=8 supergravity to an eleven dimensional setting. It consists of an M2 brane delocalised within an M5 brane, the mass of the solution is proportional to $\sqrt{Q^2+P^2}$, where $Q$ is the (electric) charge of the M2 brane and $P$ the (magnetic) charge of the M5 brane. The mass being proportional to the square root (or sum of square roots) of the charges squared is a characteristic feature of non-marginal, bound states and indicative of the mass energy of the individual states contributing to a binding energy. The dyonic membrane preserves half of the supersymmetries of the background and indeed half-BPS, non-marginal states in general, are characterised by one harmonic function and a number of parameters which interpolate between the constituent marginal brane solutions. In the dyonic membrane solution there is a parameter which interpolates between the M2 and the M5 brane solutions. In the type II supergravites other half-supersymmetric bound states include the $SL(2,Z)$ multiplet of dyonic bound states of the F1 string and the D1 brane \cite{Schwarz:1995p2725} and bound states of strings and Dp-branes \cite{Witten:1995p2727}.

There are a number of techniques used to construct bound states of M-branes. The most common of these makes use of U-duality, but solutions have also been constructed in great generality by an analysis of zero modes \cite{Adawi:1999p2616,Cederwall:1999p2334,Cederwall:2000p2304}. Such solutions are interesting for a number of reasons beyond cataloguing the non-perturbative, strongly coupled regime of M-theory however we will study their existence in this paper for their own intrinsic value. There is no systematic method for constructing non-marginal solutions from marginal solutions in eleven dimensions. It will be shown in this paper that M-theory bound state solutions can be extracted from the Kac-Moody algebra $E_{11}$ using a straightforward ansatz to relate the bound states to a number of marginal solutions.

The Kac-Moody algebra $E_{11}$ is conjectured to be a symmetry algebra of M-theory \cite{West:2001p131} and an ansatz has been found, in the form of a solution-generating group element, which encodes the half BPS M-brane solutions \cite{West:2004p1593}\footnote{The KK-monopole was not included in the solutions of \cite{West:2004p1593}, although one can see from \cite{Englert:2007p605} where the solution is presented in the context of $E(11)$ that it also fits into the solution generating group element ansatz. We will review this solution later in this paper.}. To each positive root of the adjoint representation of $E_{11}$ the method related a half BPS M-brane solution and a harmonic function. Since there are an infinite number of such roots the construction suggested there existed an infinite set of unknown, exotic branes beyond the four basic M-branes. These exotic branes would also be described in terms of one harmonic function and would preserve, supposedly, one half of the supersymmetry. Upon dimensional reduction the mixed symmetry tensors associated to exotic brane solutions are of direct use for deriving maximal gauged supergravity theories \cite{Riccioni:2007p609,Riccioni:2008p548} in lower dimensions, as ten-forms used to complete the IIB superalgebra \cite{Bergshoeff:2005p935,West:2006p77} or as gauge potentials for Dp-branes in ten dimensions \cite{Kleinschmidt:2004p371}. However there is some doubt about the role played by the full $E_{11}$ symmetry in M-theory and especially whether the high level roots associated to exotic branes have a physical significance directly in eleven dimensions. Work on U-duality multiplets of brane charges in lower dimensions \cite{Lu:1996p221,Elitzur:1998p319,Obers:1998p96,Obers:1998p275,Obers:1999p939} revealed the existence of exotic brane charges and also their need for a higher dimensional origin. In \cite{West:2004p573,Cook:2008p936} it was shown that the exotic brane charges could be derived from the fundamental, or $l_1$, representation of $E_{11}$. Furthermore the tensions of these exotic solutions were derived and it was seen that with the exception of the KK-wave, the M2 brane and the M5 brane all the branes related to positive roots within the $E_{11}$ adjoint representation had divergent tensions in a non-compact background, indicative of solutions in the strongly coupled regime. This observation is a clue to interpret such branes as non-marginal bound state solutions which is the purpose of this paper. A second clue lies in the observation that the tensions of the exotic states indicate objects which sweep out more dimensions with their worldvolume than there are dimensions in the background spacetime. A simple resolution to this problem occurs if the exotic states are actually composite solutions.

In section 2 we demonstrate our preferred method for visualising the generators of the $E_{11}$ algebra as tensor representations of $SL(11)$. By dint of the choice of vector space basis and inner product a very rapid, back-of-the-envelope generation of the Young tableaux appearing at general level, l, in the decomposition is achieved. In section 3 the half-BPS M-branes and their relation to roots of $E_{11}$ is recalled, with particular focus on the KK-monopole. In section 4 the dyonic membrane, the prototype bound state of M-branes, is studied and the relationship between the roots of $E_{11}$ and non-marginal bound state solutions is given. The central thesis of this paper is outlined: that to each positive root of $E{11}$ a non-marginal, bound state solution may be associated. In particular the two-parameter group element relating three fundamental brane solutions associated to roots: $\beta_1$, $\beta_2$ and $\beta_3\equiv \beta_1+\beta_2$ is:
\begin{align}
g_{\beta_3}=&\exp{(-\sum_i\frac{1}{\beta_i^2}\ln{N_i}{(\beta_i\cdot H})})\exp{((1-N_1)N_2^\frac{\beta_1\cdot \beta_2}{2}\sin{\xi}E_{\beta_1}}\label{interpolatinggroupelement}\\
\nonumber &+i(1-N_2)N_1^\frac{\beta_1\cdot \beta_2}{2}\tan{\xi}E_{\beta_2}+(1-N_1)N_1^{\frac{\beta_1\cdot \beta_3}{2}}N_2^{-\frac{\beta_2\cdot \beta_3}{2}}\cos{\xi}E_{\beta_3})
\end{align}
Where $N_1$ and $N_2$ are harmonic functions on the overall transverse space coordinates for the brane solutions associated to the roots $\beta_1$ and $\beta_2$. They are related by a continuous parameter, $\xi$, defined by:
\begin{equation}
\cos^2{\xi}=\frac{1-N_2}{1-N_1}
\end{equation}
We generate a number of solutions to indicate the general method including the transversely boosted M2 and M5 branes. In section 5 we consider some solutions which interpolate between three brane charges and develop the ansatz for the group element, finally using the group element to construct a bound state solution interpolating between the membrane, two fivebranes and the KK-monopole described by one harmonic function and two angle parameters.
\end{section}
\begin{section}{The $E_{11}$ algebra as Young tableaux}
Our aim in this section is to give the definitions of the Kac-Moody algebra $E_{11}$ and to derive the algebraic content of its adjoint representation in a simple way as products of $SL(11)$ Young tableaux using the Littlewood-Richardson rule.

The Kac-Moody algebra of $E_{11}$ is defined from its Cartan matrix, $A_{ab}$, which may be read off from its Dynkin diagram which is shown in figure \ref{E11}.
\begin{figure}[h]
\centering
\includegraphics[scale=0.8,angle=0]{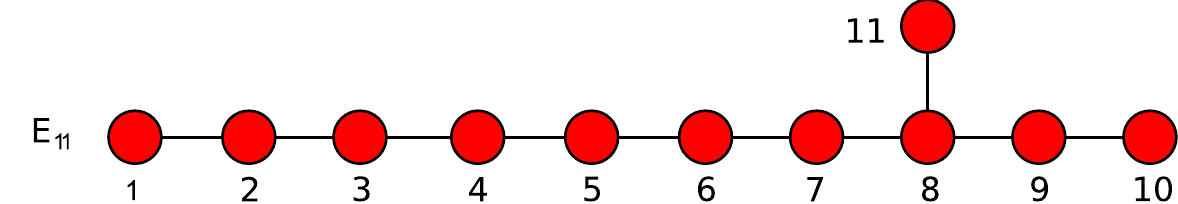}
\caption{The Dynkin diagram of $E_{11}$} \label{E11}
\end{figure}
There is a node on the Dynkin diagram for each positive, simple root, $\alpha_a$, and the Cartan matrix is defined in terms of these as:
\begin{equation}
A_{ab}=2\frac{<\alpha_a,\alpha_b>}{<\alpha_a,\alpha_a>}
\end{equation}
The Kac-Moody algebra is defined in terms of its Cartan matrix as:
\begin{align}
&[H_a,E_b]=A_{ab}E_b\qquad [H_a,F_b]=-A_{ab}F_b \qquad [E_a,F_b]=\delta_{ab}H_a\\
&[E_a,[E_a,\ldots [E_a,E_b]\ldots ]]=0 \qquad [F_a,[F_a,\ldots [F_a,F_b]\ldots ]]=0 \label{Serre}
\end{align}
Where $H_a$, $E_a$ and $F_a$ are the Chevalley generators of the Cartan subalgebra, the positive roots and the negative roots respectively. In each of the relations of the the second line, (\ref{Serre}), there are $(1-A_{ab})$ commutators - these relations are the Serre relations. The relations of the first line are easily understood as r copies of the $SU(2)$ algebra, where r is the rank of the Cartan matrix. In the Dynkin diagram there are r nodes and each node indicates an $SU(2)$ algebra. The number of lines connecting the nodes a and b in the Dynkin diagram is given by the negative off-diagonal entries of the Cartan matrix, namely $-A_{ab}$. When two positive root generators do not commute, there exists a third positive root generator whose associated root is the sum of the two roots associated to the two generators in the commutator:
\begin{equation}
[E_a,E_b]=E_{a+b}\neq 0 \qquad \Leftrightarrow \qquad \alpha_a+\alpha_b\in \Pi^+
\end{equation} 
Where $\Pi^+$ is the set of positive roots of the algebra. The Serre relations indicate the termination of a root string in the algebra, but being a set of nested commutators are very difficult to work with. Some of the information encoded in the Serre relations can be expressed as a condition on the root length squared, that, if met, indicates the existence of that root in the root system and an associated generator in the algebra. Some information is lost; we will not recover the root multiplicity in this fashion - however this can be calculated recursively using, for example, the Peterson formula. Let us derive the condition on the root length squared from the Serre relations for $E_{11}$.

The algebra $E_{11}$ is a simply-laced algebra, meaning that all its simple roots have the same length. We will normalise our roots so that their length squared is two. This simplifies the definition of the Cartan matrix:
\begin{equation}
\alpha_a^2\equiv<\alpha_a,\alpha_a>=2 \qquad \Leftrightarrow \qquad A_{ab}=<\alpha_a,\alpha_b>
\end{equation}
For a simply laced algebra, the entries of the Cartan matrix can take only three values: 2 for $A_{aa}$; -1 for $A_{ab}$ if node $a$ is connected to node $b$ in the Dynkin diagram; and 0 for $A_{ab}$ if nodes $a$ and $b$ are not directly connected by a line on the Dynkin diagram. This gives three possible expression for the Serre relations on the simple positive root generators\footnote{It is sufficient to focus on the positve root generators, since the negative root generators will have an associated root which is simply the negative of its positive counterpart. If the positive root exists, then so does the negative root.} summarised in table \ref{Serretable}.
\begin{table}[ht]
\centering
  \begin{tabular}{ | c | c | c | }
   \hline
    $A_{ab}$&$1-A_{ab}$&Serre Relation \\
    \hline
    2 & -1 & $[E_a,E_a]=0$ \\ 
    \hline
    0 & 1 & $[E_a,E_b]=0$ \\ 
    \hline
    -1 & 2 & $[E_a,[E_a,E_b]]=0$ \\
    \hline
  \end{tabular}
 \caption{The Serre relations on the generators of the positive simple roots of a simply-laced algebra} \label{Serretable}
\end{table}
The first two cases are not interesting, but in the third case the commutator $[E_a,E_b]$ may be non-zero - this is the case when the nodes $a$ and $b$ are connected by a line on the Dynkin diagram. In the case of $E_{11}$ if we commence with the root $\alpha_{11}$ we may construct a new root $\alpha_{11}+\alpha_8$, since there is a connection between nodes 11 and 8 on the Dynkin diagram. We could not, for example, find a root $\alpha_{11}+\alpha_{10}$ since nodes 11 and 10 are not connected on the Dynkin diagram. The root system of a simply-laced algebra consists of sums of simple roots which are all connected when superposed on the Dynkin diagram. In this way the Serre relation imposes irreducibility on any representation of the generators of the positive roots. Conversely if $\alpha_{11}+\alpha_{10}$ had been a root then the representation of the generators would have been reducible since there would be a subrepresentation of $SU(2)$ associated to node ten of the Dynkin diagram. We have sketched a direct relation between $A_{ab}=<\alpha_a,\alpha_b>$ being negative and the Serre relations giving a non-zero commutator for the positive root generators. Consequently we can now relate the Serre relations to a condition on the root length.

Consider the root length for the case we have just considered of adding two simple roots, $\alpha_a$ and $\alpha_b$, to find a new root $\alpha_a+\alpha_b$. In this case the new root length is:
\begin{align*}
(\alpha_a+\alpha_b)^2&=\alpha_a^2+\alpha_b^2+2<\alpha_a,\alpha_b>\\
&=2+2+2A_{ab}\\
&=2
\end{align*}
Where we have used the observation that if $\alpha_a+\alpha_b$ is a root then $A_{ab}=-1$. Now suppose we have a third root, $\alpha_c+\alpha_b+\alpha_a$, where $\alpha_c$ is another simple root. Since we have asserted it is a root we see that
\begin{equation}
[E_c,E_{a+b}]=[E_c,[E_a,E_b]]=E_{a+b+c}\neq 0
\end{equation}
Let us expand the nested commutator expression using the Jacobi Identity:
\begin{equation}
[E_c,[E_a,E_b]]=[[E_c,E_a],E_b]+[E_a,[E_c,E_b]]\neq 0
\end{equation}
Both the left hand side and the right hand side are non-zero, so we have three possible conditions from the Serre relations: 
\begin{enumerate}
\item $A_{ab}=-1$, $A_{ac}=-1$ and $A_{bc}=0$
\item $A_{ab}=-1$, $A_{ac}=0$ and $A_{bc}=-1$
\item $A_{ab}=-1$, $A_{ac}=-1$ and $A_{bc}=-1$
\end{enumerate}
Let us consider the root length squared for the different cases:
\begin{align*}
(\alpha_a+\alpha_b+\alpha_c)^2&=\alpha_a^2+\alpha_b^2+\alpha_c^2+2<\alpha_a,\alpha_b>+2<\alpha_a,\alpha_c>+2<\alpha_b,\alpha_c>\\
&=2+2+2+2A_{ab}+2A_{ac}+2A_{bc}\\
&=\left\{
\begin{array}{lr}
2&\text{cases 1 and 2} \\
0&\text{case 3\hspace{34pt}} 
\end{array}
\right.
\end{align*}
Cases 1 and 2 are the minimal examples for the existence of a new root according to the Serre relations, in these cases node $c$ is connected to only one of nodes $a$ and $b$ and the root length squared is equal to two. In case 3 node $c$ is connected to both nodes $a$ and $b$ and the root length squared decreases by two. One may continue this process and build up the root string, it is clear that if one does this that the Serre relations provide a bound on the root length squared. Explicitly, a root, $\beta$, in a simply-laced algebra will always satisfy:
\begin{equation}
\beta^2=2,0,-2,-4 \ldots
\end{equation}
Our intention now is to apply this condition to the roots associated to generators with the correct index structure to appear in the algebra. 

By the level decomposition of the $E_{11}$ algebra into $SL(11)$ representations the generators of $E_{11}$ may be represented by Young tableaux. Recall that upon deletion of the eleventh node of the $E_{11}$ Dynkin diagram the remaining diagram is that of $A_{10}$, or $SL(11)$. A generic root associated to the $E_{11}$ algebra requires multiple deletions of $\alpha_{11}$ before a root of $SL(11)$ remains, the number of deletions of $\alpha_{11}$ needed for this to occur is called the level of the decomposition. The deleted root can be written as a sum of two parts, the first part, $x$, being orthogonal to all the simple roots of $A_{10}$ and the second part, $-\lambda_8$, being in the $A_{10}$ weight lattice:
\begin{equation}
\alpha_{11}=x-\lambda_8
\end{equation}
Where $\lambda_8$ is a fundamental weight of $A_{10}$, having the defining property that $<\lambda_i,\alpha_j>=\delta_{ij}$. Consequently the inner products between the roots of $A_{10}$ and $\alpha_{11}$ are those of $E_{11}$ Cartan matrix. The deletion of $\alpha_{11}$ corresponds to a representation of $A_{10}$ whose highest weight is $\lambda_8$ which is a rank three antisymmetric $SL(11)$ tensor. 

At this point it will be useful to express the roots of $E_{11}$ in a vector space basis, $\{e_1,e_2,\ldots e_{11}\}$, instead of a simple root basis, $\{\alpha_1,\alpha_2,\ldots \alpha_{11}\}$. One can rewrite the simple roots as:
\begin{align}
\alpha_i&=e_i-e_{i+1} \qquad \qquad i\leq 10\\
\nonumber \alpha_{11}&=e_9+e_{10}+e_{11}
\end{align}
On a vector space endowed with inner product:
\begin{equation}
<\vec{a},\vec{b}>=\sum_{i=1}^{11}a_ib_i-\frac{1}{9}\sum_{i=1}^{11}a_i\sum_{j=1}^{11}b_j \label{innerproduct}
\end{equation}
Where,
\begin{equation}
\vec{a}=\sum_i a_i e_i, \qquad \vec{b}=\sum_i b_i e_i
\end{equation}
One can check that the roots expressed in this basis with this inner product obey the inner products of the $E_{11}$ Cartan matrix. Now we are in a position to notice a curious thing, namely that in this basis the index structure of the associated generators may be read off immediately from the root.  The rule to follow is simply that $+e_i$ indicates a contravariant index (carrying the label $i$ of the vector $e_i$), while $-e_i$ indicates a covariant index and that if a coefficient is greater than one these indices are symmetrised. For example consider the roots $\{\alpha_1,\ldots, \alpha_{10}\}$ which are the simple positive roots of $A_{10}$. Following these rules we can write down the simple root generators of $A_{10}$ ($\alpha_i=e_i-e_{i+1}$) as ${K^i}_{i+1}$ by simply reading off the coefficients of the roots in the $e_i$ basis. For $\alpha_{11}=e_9+e_{10}+e_{11}$ we write down the generator $R^{91011}$, and for $\beta_{M5}=e_6+e_7+\ldots+e_{11}$ we find the generator $R^{67\ldots 11}$. Since the full $E_{11}$ algebra is constructed from multiple commutators of these generators the rule for reading off the index structure of the $SL(11)$ tensor generators from the root vector in the $\{e_i\}$ basis is true for the full algebra. As an example of a mixed symmetry generator consider the KK-monopole whose gauge field is associated to the generator $R^{4\ldots 11, 11}$ - one readily deduces that the associated root is $e_4+\ldots +e_{10}+2e_{11}$.  For a general root $\beta\equiv \sum_i c_i e_i$ the Young tableau for the associated generator has $c_{11}$ boxes in its top row, $c_{10}$ boxes in the next row and so on down to $c_1$ boxes in the bottom row. We are able to rapidly associate Young tableaux to roots of $E_{11}$ expressed in the $e_i$ basis or vice-versa. 

Returning to the decomposition and having observed that the deleted root, $\alpha_{11}=e_9+e_{10}+e_{11}$, is associated an antisymmetric three tensor we can construct the $E_{11}$ algebra. At level one we draw the Young tableau for an antisymmetric 3-tensor:
\begin{equation}
\yng(1,1,1)
\end{equation}
To be explicit we would write the coordinates 11, 10, 9 descending from top to bottom in the Young tableau, this would correspond to a highest weight $SL(11)$ tensor. In the following we will assume an empty Young tableau indicates the highest weight $SL(11)$ tensor: each column will be associated to coordinates starting with 11 in the top box and decreasing in units until the base of each column is reached. We may now check the root length of the Young tableau (to confirm that it is not projected out by the Serre relations) using (\ref{innerproduct}), and find it is two, as expected and it is in the root system.
At the next level we use the Littlewood-Richardson rule to construct the products of $\tiny \yng(1,1,1)$ with the Young tableaux associated to roots at the previous level. At level two we find the highest weight Young tableaux:
\begin{equation}
\Yvcentermath1
\yng(1,1,1) \otimes \yng(1,1,1)=\yng(1,1,1,1,1,1) \oplus \yng(2,1,1,1,1)\oplus \yng(2,2,1,1)\oplus \yng(2,2,2)
\end{equation}
These roots have lengths 2, 4, 6 and 8 respectively. Hence only the first Young tableau, corresponding to an antisymmetric 6-tensor, appears at level 2.
At level three we have:
\begin{equation}
\Yvcentermath1
\yng(1,1,1,1,1,1) \otimes \yng(1,1,1)=\yng(1,1,1,1,1,1,1,1,1) \oplus \yng(2,1,1,1,1,1,1,1)\oplus \ldots
\end{equation}
The Young tableaux shown have roots with length squared 0 and 2, and the rest of the Young tableaux do not appear in the algebra because their associated root length squared is greater than two. It is straightforward to check that the consequence of moving a box from one column to an adjacent one directly to its right, while still having a valid Young tableau (i.e. the columns have equal or descending length from left to right) is to add two to the root length squared. Consequently the root length condition means one constructs only the tallest and thinnest Young tableaux at each level. One can give a general rule for constructing the highest weight Young tableaux at an arbitrary level, l, associated to the $E_{11}$ algebra: take 3l boxes and arrange them into the tallest and thinnest Young tableaux, with the constraint that the column height is at most eleven. Those that have length squared less than or equal to two are, up to multiplicity considerations, associated to generators present in the algebra. At level three, considered above, the completely antisymmetric 9-tensor is not present in the algebra due to multiplicity considerations. It appears there is at least one Young tableau per level, after level two, which has outer multiplicity of zero and hence is not present in the algebra \cite{Nicolai:2002p1596}. It would be interesting to understand the relatively infrequent occurence of roots with outer multiplicity zero. For the purpose of this paper we will ignore multiplicity considerations.

\end{section}
\begin{section}{$E_{11}$ and the standard M-theory solutions}
A solution-generating group element encoding $\frac{1}{2}$-BPS solutions of eleven dimensional supergravity was found in \cite{West:2004p1593} and its generalisation in \cite{Cook:2005p270} was shown to reconstruct all the $\frac{1}{2}$-BPS solutions to the maximally oxidised supergravities associated to a $\cal{G}^{+++}$ symmetry.
The group element takes the following form
\begin{equation}
g_{\beta}=\exp{(-\frac{1}{\beta^2}\ln{N} H\cdot\beta)}\exp{((1-N)E_\beta)} \label{halfbpsgroupelement}
\end{equation}
Where $\beta$ is a root in the adjoint representation of $E_{11}$, $E_\beta$ is its associated generator in the algebra, $H$ are the elements of the Cartan sub-algebra and $N$ is a harmonic function. The harmonic function $N$ appears as a solution to the equations of motion found by varying the generic gravity action,
\begin{equation}
{\cal{L}}=\int d^{11}x \sqrt{-g}\{R-\frac{1}{n!}(F^{a_1\ldots a_n})^2\}
\end{equation}
Where $F^{a_1\dots a_n}$ is the field strength derived from the potential $A_{a_1\ldots a_{n-1}}$. This gauge field appears as the coefficient of the generator $E_\beta$ in the group element of equation (\ref{halfbpsgroupelement}). As such the group element gives an on-shell description of the $\frac{1}{2}$-BPS brane solutions. Note that for these solutions described by a single active potential the Chern-Simons term vanishes, for more complicated potentials this will no longer be the case.

\begin{subsection}{Fundamental solutions of M-theory}
Let us familiarise ourselves with the group element by reviewing the examples of the KK-wave, the M2-brane  and the M5-brane solutions as given in \cite{West:2004p1593}. The KK-monopole, which we will refer to herein as the KK6-brane or KK6-monopole\footnote{The gauge field is also known as the dual gravity field and is a purely gravitational solution.} as upon dimensional reduction it gives rise to the D6 brane has been related to a group element of the form of (\ref{halfbpsgroupelement}) \cite{Englert:2007p605}.  Here we will review how it is encoded in the group element but in a more complicated way than the other half-BPS solutions. 

The KK-wave solution is associated to the roots of the $SL(11)$ sub-algebra which appear at level 0 in the decomposition of the $E_{11}$ algebra. For example consider the case when $\beta_{KK}=\alpha_{10}$ in (\ref{halfbpsgroupelement}), so that $E_{\beta}={K^{10}}_{11}$. In this case,
$$H\cdot \beta_{KK} = {K^{10}}_{10}-{K^{11}}_{11}$$
We use (\ref{halfbpsgroupelement}) to read off the line element:
$$ds^2=dx_1^2+\ldots+dx_9^2-N^{-1}dt_{10}^2+N(dx_{11}-(N^{-1}-1)dt_{10})^2$$
Where we have chosen the tenth coordinate to be timelike. Note that the gauge field in this case gives an off-diagonal vielbein component premultiplying the generator ${K^{10}}_{11}$, in tangent space indices this is just $(1-N)$ whereas in worldvolume indices it becomes $(N^{-1}-1)$, the factor rotating $dx_{11}$ into $dt_{10}$. The ansatz for solution generation is to take the function $N(x_1,\ldots, x_9)$ to be a harmonic function, specifically $N=1+\frac{Q}{r^7}$.

The M2-brane solution is associated to the simple root $\beta_{M2}=e_9+e_{10}+e_{11}$ of $E_{11}$ appearing at level one in the algebra. This root is associated to the highest weight of the generator $R^{a_1a_2a_3}$, the component $R^{91011}(\equiv E_\beta)$. Consequently,
\begin{equation}
H \cdot \beta_{M2} =-\frac{1}{3}({K^1}_1+\ldots +{K^8}_8)+\frac{2}{3}({K^9}_9+\ldots+{K^{11}}_{11})
\end{equation}
From equation (\ref{halfbpsgroupelement}) one can read off the vielbein components to find the line element of the M2 brane,
\begin{equation}
ds_{M2}^2=N^{\frac{1}{3}}(dx_1^2+\ldots dx_8^2)+N^{-\frac{2}{3}}(-dt_9^2+dx_{10}^2+dx_{11}^2)\label{M2}
\end{equation}
Where $N(x_1,\ldots x_8)$ is a harmonic function of the transverse coordinates:
\begin{equation}
N=1+\frac{Q}{r^6}
\end{equation}
It satisfies $d\star F=0$, where $F=dN^{-1}$, which can be rewritten as the curved space Laplace equation:
\begin{equation}
\partial_\mu (\sqrt{|g|}g^{\mu\nu}g^{99}g^{1010}g^{1111}F_{\nu91011})=0
\end{equation} 

The $M5$ brane is derived from the level two root $\beta_{M5}=e_6+\ldots +e_{11}$, so that,
\begin{equation}
H\cdot \beta_{M5}=-\frac{2}{3}({K^1}_1+\ldots {K^5}_5)+\frac{1}{3}({K^6}_6+\ldots + {K^{11}}_{11})
\end{equation}
and the method gives the line element:
\begin{equation}
ds_{M5}^2=M^{\frac{2}{3}}(dx_1^2+\ldots dx_5^2)+M^{-\frac{1}{3}}(-dt_6^2+\ldots+dx_{10}^2+dx_{11}^2) \label{M5}
\end{equation}
Where $M(x_1,\ldots x_5)$ is harmonic, so that:
\begin{equation}
M=1+\frac{Q}{r^3}
\end{equation}
The root at level four has $\beta_{KK6}=e_4+\ldots e_{10}+2e_{11}$ and is associated to the KK6-brane. The relation between the harmonic functions encoding the  KK6 monopole solution and the root $\beta_{KK6}$ were given in \cite{Englert:2007p605}. The solution has not been understood in terms of the solution generating group element (\ref{halfbpsgroupelement}) previously and so we discuss its derivation in detail here.

We compute:
\begin{equation}
H\cdot \beta_{KK6}=-({K^1}_1+\ldots {K^3}_3)+{K^{11}}_{11}
\end{equation}
Which gives a diagonalised metric:
\begin{equation}
ds_{KK6}^2=N(dx_1^2+\ldots dx_3^2)+N^{-1}(dx_{11}^2)^2+d\Omega_{(1,6)}^2
\end{equation}
Unlike the membrane and fivebrane solutions the KK6-brane is a pure gravity solution. Its gauge field $A^{4\ldots 11,11}$ is dual to the vielbein field and as with the $KK$-wave adds an off-diagonal contribution to the metric. The field strength associated to the dual gravity field is dualised as:
\begin{equation}
\frac{1}{9!}\epsilon^{a_1\ldots a_{11}}F_{a_3\ldots a_{11},b}={F^{a_1a_2}}_b=2\partial^{[a_1}{h^{a_2]}}_b
\end{equation}
In particular for the highest weight with component $A_{4\ldots 11,11}$, whose field strength is related to the harmonic function $N(x_1,x_2,x_3)$ as:
\begin{equation}
F_{k4\ldots 11,11}=\partial_k N^{-1}=-N^{-2}\partial_k N\quad \Rightarrow \quad F^{k4\ldots 11,11}=-N\partial^k N=-\partial_k N
\end{equation}
Where $i,j,k\in\{x_1,x_2,x_3\}$, and by raising the indices prior to taking the Hodge dual we ensure that the dual field has a set of covariant indices. Consequently,
\begin{equation}
{F_{ij}}^{11}=2\partial_{[i}{h_{j]}}^{11}=-\epsilon_{ijk}\partial_k N\equiv \vec{\nabla} \wedge {\vec{A}} \label{dualkk}
\end{equation}
From the solution generating group element we conclude that it has line element:
\begin{align}
ds_{KK6}^2&=N(dx_1^2+\ldots dx_3^2)+N^{-1}(dx_{11}^2-A_i\cdot dx_i)^2+d\Omega_{(1,6)}^2 \label{KK6}
\end{align}
Where $N(x_1,x_2,x_3)=1+\frac{2K}{r}$ is a harmonic function and $i=\{1,2,3\}$. This is a Wick rotation of the Gross-Sorkin-Perry KK-monopole. Consider as a particular example that case when only the off-diagonal veilbein component ${h_3}^{11}$ is non-zero, which corresponds to limiting $N$ to be harmonic in only $x_1$ and $x_2$ \cite{Englert:2007p605}, then the line element is a smeared gravitational monopole:
\begin{align}
\nonumber ds_{KK6}^2&=N(dx_1^2+\ldots dx_3^2)+N^{-1}(dx_{11}^2+(1-N)dx_3)^2-dt_4^2+\ldots+dx_{10}^2 \\
N(x_1,x_2)&=Q\ln(r)
\end{align}
Where we have used equation (\ref{dualkk}) to identify the components $A_i$ with the field ${h_{3}}^{11}=1-N$.

Returning to the general KK6 brane solution of equation (\ref{KK6}) and rewriting the metric in spherical coordinates, $\theta, \phi$ and a shifted radial coordinate, $\hat{r}=r+K$ we arrive at a form derived from the Euclidean Taub-NUT metric embedded in eleven dimensional space-time:
\begin{align}
\nonumber ds_{NUT}^2=&N(d{r}^2)+(\hat{r}^2-K^2)(d\theta^2+\sin^2{\theta}d\phi^2)+N^{-1}(dx_{11}^2-2K\cos{\theta}d\phi)^2\\
\nonumber &+d\Sigma^2_{1,6}\\
N=&\frac{\hat{r}+K}{\hat{r}-K}
\end{align}
This is the extremal version of the Euclidean Taub-NUT metric in eleven dimensions. 

The new technique apparent in deriving this solution from the roots of $E_{11}$ is the use of dual gauge potentials. As we see above choosing to work with a dual field can cast a solution in a recognisable form. In the case of the KK6-brane the $(9,1)$ field strength that one expects to be associated to the $(8,1)$ gauge field is recast as the field strength of the dual graviton and is absorbed into the volume element.
\begin{subsubsection}{Hodge Duality}
The fundamental solutions are paired up by Hodge duality, the $M2(S2)$ brane is related to the $S5(M5)$ brane and the KK wave to the KK6 monopole. In terms of the roots that the solutions are derived from the dual solutions are mapped to each other by:
\begin{equation}
\beta_{*Mp}=\beta_{9}-\beta_{Mp}
\end{equation} 
Where:
\begin{equation}
\beta_{9}\equiv e_3+e_4+\ldots + e_{11}
\end{equation}
This is a root at level three in the decomposition of $E_{11}$ into representations of $SL(11)$ which does not appear in the algebra due to multiplicity considerations. It is associated to a tensor with nine antisymmetrised indices. Hodge duality in eleven dimensions relates gauge fields with $p+1$ antisymmetric indices to those with $9-(p+1)$ antisymmetric indices. Hence the role of $\beta_9$ in relating dual solutions. We propose the following group element for dual solutions:
\begin{align}
\nonumber g_\beta&=\exp{(-\frac{1}{\beta_{*Mp}^2}\ln N (H\cdot \beta_9)+\frac{1}{\beta_{*Mp}^2}\ln N (H\cdot \beta_{Mp}))}\\
&\qquad \exp{((1-N)N^{\frac{-\beta_{Mp}\cdot \beta_{*Mp}}{2}}E_{\beta_{*Mp}})} \label{dualgroupelement}
\end{align}
Consider the example of the $S5$ derived from the $M2$ group element. Taking,
\begin{align}
\beta_{M2}&=e_9+e_{10}+e_{11}, \qquad \beta_{S5}=\beta_9-\beta_{M2}=e_3+\ldots +e_8\\
\nonumber & \qquad \qquad \Rightarrow \beta_{S5}\cdot\beta_{M2}=-2
\end{align}
Substituting this into \ref{dualgroupelement} gives:
\begin{align}
\nonumber g_\beta&=\exp{(-\frac{1}{2}\ln N (H\cdot \beta_{S5}))}\exp{((1-N)NR^{91011})}
\end{align}
Reading off the metric gives:
\begin{equation}
ds^2=N^{\frac{2}{3}}(N^{-1}(dx_3^2+\ldots dx_8^2)+dx_1^2+dx_2^2+dx_9^2+\ldots dx_{11}^2)
\end{equation}
Using the vielbein to convert the tangent space indices of the gauge field to worldvolume indices we find the dual field strength:
\begin{equation}
{\cal F}_4=-dN
\end{equation}
Similarly we can directly find the KK6 monopole metric (up to a Wick rotation) from the KK-wave group element. The relevant roots and inner products are:
\begin{align}
\beta_{KK}&=e_3-e_{11}, \qquad \beta_{KK6}=\beta_9-\beta_{KK}\\
\nonumber & \qquad \Rightarrow \beta_{KK}\cdot \beta_{KK6}=-2
\end{align}
The dualised group element for the KK6 monopole is:
\begin{align}
g_\beta&=\exp{(-\frac{1}{2}\ln N (H\cdot \beta_{\beta_{KK9}}))}\exp{((1-N)N{K^3}_{11})} \label{KK6dualgroupelement}
\end{align}
After converting the gauge field to worldvolume indices one reproduces the KK6 monopole volume element, and one can see clearly that the solution is purely gravitational in this dual form. 
\end{subsubsection}
\end{subsection}

\begin{subsection}{Marginal intersecting M-brane solutions}
The intersecting brane solutions were first understood in \cite{Englert:2004p2875,Englert:2004p2874} to arise for branes associated to roots, $\beta_1$ and $\beta_2$, such that $\beta_1\cdot \beta_2=0$. The combination of two solution generating group elements (\ref{halfbpsgroupelement}) one finds is \cite{West:2004p1593}:
\begin{align}
\nonumber g_1g_2&=\exp{(-\frac{1}{\beta^2}\ln{N_1}(H\cdot \beta_1)-\frac{1}{\beta^2}\ln{N_2}(H\cdot \beta_2))}\\
&\exp{((1-N_1)N_2^{-\frac{\beta_1\cdot\beta_2}{2}}E_{\beta_1})}\exp{((1-N_2)E_{\beta_2})} \label{marginalgroupelement}
\end{align}
This group element does not possess a manifest symmetry between the harmonic functions of the two brane solutions, except in the case when $\beta_1\cdot \beta_2=0$. The cases when this occurs correspond to the basic M-brane intersections. We show examples of the roots and the associated marginal solutions in table \ref{marginal}.
\begin{table}[ht]
\centering
  \begin{tabular}{ | c | c | c | }
   \hline
    $\beta_1$&$\beta_2$&Marginal solution \\
    \hline
    $e_{10}-e_{11}$ & $e_9+e_{10}+e_{11}$ & Boosted $M2$ \\
    \hline
    $e_9+e_{10}+e_{11} $ & $e_7+e_8+e_{11}$ & $M2\perp M2(0)$ \\ 
    \hline
    $e_{10}-e_{11}$ & $e_6+e_{10}+e_{11}$ & Boosted $M5$ \\
    \hline
    $e_9+e_{10}+e_{11}$ & $e_6+\ldots+e_{11}$ & $M2\perp M5(1)$ \\
    \hline 
    $e_6+\dots+e_{11}$& $e_4+e_5+e_8+\ldots+e_{11}$ & $M5\perp M5(3)$ \\ 
    \hline
    $e_{9}-e_{10}$ & $e_4+\ldots +e_{10}+2e_{11}$ & Boosted $KK6$ \\
    \hline
    $e_4+\ldots +e_{10}+2e_{11}$ & $e_3+e_4+e_{11}$ & $KK6\perp M2(1')$ \\
    \hline
    $e_4+e_5+e_6$& $e_4+\ldots +e_{10}+2e_{11}$ & $M2\in KK6$ \\
    \hline
    $e_4+\ldots +e_{10}+2e_{11}$ & $e_3+e_4+\ldots +e_7+e_{11}$ & $KK6\perp M5(4')$ \\
    \hline
    $e_4+\ldots+e_9$ & $e_4+\ldots +e_{10}+2e_{11}$& $M5\in KK6$ \\
    \hline
    $e_4+\ldots +e_{10}+2e_{11}$ & $e_2+e_3+e_4+\ldots +e_{8}+2e_{11}$ & $KK6 \perp KK6 (5'')$ \\
    \hline
    $e_4+\ldots +e_{10}+2e_{11}$ & $e_3+e_4+\ldots +e_{8}+2e_{10}+e_{11}$ & $KK6 \perp KK6 (6'')$ \\
    \hline
     $e_9+e_{10}+e_{11}$ & $e_7+e_8+e_9$ & $M2\perp S2(1)$ \\
    \hline
    $e_9+e_{10}+e_{11}$ & $e_5+\ldots e_{10}$ & $M2\perp S5(2)$ \\
    \hline
    $e_6+\ldots e_{11}$ & $e_4+\ldots e_9$ & $M5\perp S5(4)$ \\
    \hline
  \end{tabular}
 \caption{Roots satisfying $\beta_1\cdot \beta_2=0$ associated to basic marginal brane intersections.} \label{marginal}
\end{table}

In table \ref{marginal} the number in brackets indicates the number of overlapping spacelike dimensions and we indicate that the Taub-NUT direction of the KK6-brane is included amongst the intersection directions by a prime, so that $KK6\perp M2(1')$ indicates a KK-monopole intersecting an M2 brane over a string in the Taub-NUT direction. These solutions, and mutliple combinations of them, were originally found using properties of supersymmetry, dimensional reduction, the method of harmonic superposition (subsequently identified as a no-force condition between the constituent branes) and analysis of the equations of motion \cite{Papadopoulos:1996p1472,Tseytlin:1996p1469,Gauntlett:1996p1657,Tseytlin:1996p1683,Arefeva:1996p2729,Khviengia:1996p2040,Lu:1996p115,Costa:1996p1651,Argurio:1997p1652,Bergshoeff:1997p694,Deger:2002p1845,Ohta:2003p1840}. The solutions in table \ref{marginal} are described by two harmonic functions and preserve $\frac{1}{4}$ of the supersymmetry of the background. The solutions may be treated as building blocks and one may combine the intersection rules to find solutions with $N$ harmonic functions and preserving $\frac{1}{2^N}$ of the supersymmetries.  For example one can construct a $\frac{1}{8}$-BPS solution involving three M2 branes each of which intersects the other M2 branes over a point, i.e. $M2(012)\perp M2(034)\perp M2(056)$. The limiting factor in constructing these intersections is the dimension of the background spacetime, or, equivalently, the amount of supersymmetry that may be broken.

For any of the solutions of table \ref{marginal} the volume element may be reconstructed using the harmonic superposition rule \cite{Tseytlin:1996p1469}. The two metrics of the contributing brane solutions are superposed and their harmonic functions are restricted to be harmonic in only the overall transverse directions. This is equivalent to smearing a constituent brane over the worldvolume directions of the other constituent branes. For example consider the $M2\perp M2 (0)$ solution, from the solution generating group element for $\beta_1$ and $\beta_2$ (which we can read from table \ref{marginal}), the two membranes have the following worldvolumes:
\begin{align}
ds_1^2&=N_1^{\frac{1}{3}}(N_1^{-1}(dx_9^2+dx_{10}^2-dt_{11}^2)+dy_idy^i)\qquad i\in \{1,\ldots, 8\}\\
\nonumber ds_2^2&=N_2^{\frac{1}{3}}(N_2^{-1}(dx_7^2+dx_8^2-dt_{11}^2)+dy_jdy^j)\qquad j\in \{1,\ldots, 6,9,10\}\label{2M2s}
\end{align}
Superposing the solutions gives:
\begin{align}
\nonumber ds_{1\otimes 2}^2&=N_1^{\frac{1}{3}}N_2^{\frac{1}{3}}(N_2^{-1}(dx_7^2+dx_8^2)+N_1^{-1}(dx_9^2+dx_{10}^2)-N_1^{-1}N_2^{-1}dt_{11}^2\\
&\qquad \qquad \qquad +dy_kdy^k)\qquad k\in \{1,\ldots, 6\}
\end{align}
The solutions are smeared in their relative transverse directions so that,
\begin{equation}
N_i=1+\frac{Q_i}{r^4} \qquad \mbox{where} \quad r^2=y_ky^k
\end{equation}
The field strength for the intersecting solution is a sum of the field strengths for the individual solutions:
\begin{equation}
{\cal F}_{4(2\perp2)}=-dt^{11}\wedge(dN_1^{-1}\wedge dx^9\wedge dx^{10}+dN_2^{-1}\wedge dx^7\wedge dx^8)
\end{equation}
This may be read from the group element \ref{marginalgroupelement} after the tangent space indices of the gauge fields are converted into worldspace indices under the action of the vielbein of the \emph{individual brane backgrounds} \ref{2M2s}. Finally in combining the final two exponentials of \ref{marginalgroupelement} to read off the gauge field components we note that $[E_{\beta_1},E_{\beta_2}]=0$ which will be in contrast to the non-marginal bound states we consider next.
\end{subsection}
\end{section}
\begin{section}{Composite gravitational solutions}
In this section we outline the procedure that will allow us to associate roots with non-marginal, bound states of marginal solutions. The first example we will reproduce is the dyonic membrane which we will associate to the level two root in the decomposition of $E_{11}$.

The solution generating group element is formed of two parts, the first exponential contains the gravitational solution and the second the gauge fields. The root at a given level, $l$, may always be re-expressed as a sum of roots from lower levels, $l_i$ such that: $l=\sum_{i=0}l_i$. The level 0 roots are some element of the root system of $SU(11)$. The lowest level roots, at $l=1,2,3$, correspond to well-known solutions of supergravity derived in the previous section. Given an arbitrary high level (greater than level two) root we may always partition it into a sum of roots from levels $l=0,1,2,3$ whose solutions are well understood. In this paper we will consider the decomposition of real roots but the interpretation of roots as composite solutions presented here will extend also to the null and imaginary roots, indeed it has previously been proposed \cite{Brown:2004p2806}\footnote{We thank Axel Kleinschmidt for drawing our attention to this paper.} that the null and imaginary roots of $E_{10}$ correspond to Minkowski branes, and furthermore that certain imaginary roots may be decomposed into sums of two roots in order to better investigate their physical nature in broad agreement with the central thesis of this paper.

As we will see this will result in introducing multiple parameters to describe a single solution with one harmonic function. The construction of solution generating group elements will vary from solution to solution, in this section we will propose group elements for  two parameter half-BPS bound state solutions. The construction in general may be complicated by the large number of parameters involved but the steps taken here for the two parameter examples are reproducible in the more general examples which we consider in section 5.
The prototype solution that we will derive using the solution generating group element is the dyonic membrane \cite{Izquierdo:1995p1636}. 
\begin{subsection}{The dyonic membrane}
The M5 brane solution is associated to a root of $E_{11}$ that appears at level two in the decomposition. It has a unique expression as a sum of roots associated to membrane solutions:
\begin{equation}
\beta_{M5}=2\beta_{M2}+\beta_{pp}
\end{equation}
Where,
\begin{align}
\beta_{M2}&=e_9+e_{10}+e_{11}\\
\beta_{pp}&=e_6+e_7+e_8-e_9-e_{10}-e_{11}
\end{align}
The root $\beta_{pp}$ is a sum of roots in the $A_{10}$ sub-algebra of the decomposition, in fact here $\beta_{pp}=\alpha_6+2\alpha_7+3\alpha_8+2\alpha_9+\alpha_{10}$, and have the effect of rotating indices on the gauge fields. There are two gauge fields associated to the two M2 roots, $\beta_{M2}$: $R_{(1)}^{91011}$ and $R_{(2)}^{91011}$. By allowing the $SL(11)$ generators to act on the $R_{(i)}^{91011}$ generators and lower one set of the membrane indices we are left with an M2 (678) brane root, with corresponding generator $R_{(1)}^{678}$, and an S2 (91011) brane root, with generator $R_{(1)}^{678}$, where we have indicated the worldvolume directions in brackets:
\begin{equation}
\beta_{M2}=e_6+e_7+e_8, \qquad \beta_{S2}=e_9+e_{10}+e_{11}
\end{equation}
Notice that the time coordinate is rotated by the $SL(11)$ adjoint action \cite{Keurentjes:2004p1580}. Consequently we take the $x^6$ direction to be temporal in this pair of solutions. We also observe that $\beta_{M2}\cdot\beta_{S2}=-1$ which, as outlined in section 1, indicates that one can add to the root string and find a generator associated to the root $\beta_{M2}+\beta_{S2}$; this is to be contrasted with the situation for marginal brane intersections for which $\beta_1\cdot \beta_2 =0$.
Using the superposition procedure described earlier in this paper we find the volume element:
\begin{align}
\nonumber ds_{M2,S2}^2&=(N_1N_2)^{\frac{1}{3}}(dx_1^2+\ldots dx_5^2+N_1^{-1}(-dt_6^2+dx_7^2+dx_8^2)\\
&\qquad +N_2^{-1}(dx_9^2+dx_{10}^2+dx_{11}^2)) \label{dyonicmembranemetric}
\end{align}
We notice that we recover the dyonic membrane solution \cite{Izquierdo:1995p1636} if we take as our ansatz for $N_1$ and $N_2$
\begin{equation}
N_{1}=1+\frac{Q}{r^3}\quad N_{2}=1+\frac{Q\cos^2{\xi}}{r^3}
\end{equation}
Where $\xi$ is a free parameter and when $\xi=0$ ($N_2=N_1$) we have the magnetic fivebrane solution, while when $\xi=\frac{\pi}{2}$ ($N_2=1$) we have the electric membrane solution. 
This suggests the ansatz for the solution-generating group element should be augmented with an angle variable that measures the difference between the membrane charges $(1-N_1)$ and $(1-N_2)$ as
\begin{equation}
\frac{1-N_2}{1-N_1}=\cos^2{\xi} \label{ratio}
\end{equation}
To find the interpolating group element given in equation (\ref{interpolatinggroupelement}) we need the following inner products
\begin{equation}
\beta_{M2}\cdot\beta_{S2}=-1 \qquad \beta_{M2}\cdot\beta_{M5}=1\qquad \beta_{S2}\cdot\beta_{M5}=1
\end{equation}
The group element (\ref{interpolatinggroupelement}) for the dyonic membrane is:
\begin{align}
\nonumber g_{\beta}=&\exp{(-\frac{1}{\beta^2}\ln{N_1}(H\cdot \beta_{M2})-\frac{1}{\beta^2}\ln{N_2}(H\cdot \beta_{S2}))}\\
&\exp((1-N_1)^{\frac{1}{2}}(1-\frac{N_1}{N_2})^{\frac{1}{2}}R^{678}+(1-N_2)^{\frac{1}{2}}(1-\frac{N_2}{N_1})^{\frac{1}{2}}R^{91011}\label{dyonicmembranegroupelement} \\
\nonumber &\qquad +(1-N_1)^{\frac{1}{2}}(1-N_2)^{\frac{1}{2}}(\frac{N_1}{N_2})^{\frac{1}{2}}R^{6\ldots 11})
\end{align}
The solution generating group element is now dependent upon two paramaters, $r$ and $\xi$. We note that when $N_2\rightarrow1$ it approaches the group element for a membrane solution, when $N_1\rightarrow1$ it approaches the solution generating group element for a spacelike membrane solution, and when $N_1\rightarrow N_2$ it approaches the group element that encodes the fivebrane solution. Also note that under interchange of $N_1$ and $N_2$ the gauge field components for the M2 brane and the S2 brane are interchanged while the expression for the fivebrane gauge field component is mapped to itself.

We can use equation (\ref{ratio}) to write a line element in terms of $N_1(r)$ and $\xi$ giving:
\begin{align}
\nonumber ds_{M2\oplus S2}^2&=N_1^{\frac{1}{3}}(\sin^2\xi+N_1\cos^2\xi)^{\frac{1}{3}}(dx_1^2+\ldots dx_5^2+N_1^{-1}(-dt_6^2+dx_7^2+dx_8^2)\\
&\qquad +(\sin^2\xi+N_1\cos^2\xi)^{-1}(dx_9^2+dx_{10}^2+dx_{11}^2))
\end{align}
The group element in equation (\ref{dyonicmembranegroupelement}) encodes the four-form field strength of the dyonic membrane solution. It is helpful to write the gauge field for the fivebrane in a dual form, this amounts to the insertion of $N_2\frac{-(\beta_9-\beta_{M5})\cdot\beta_{M5}}{2}=N_2$ in front of the $R^{67891011}$ generator. With this insertion when $N_2=N_1$ we find the solution generating group element for an M5 brane written in a dual form. The gauge part of the group element is modified to read:
\begin{align}
\nonumber & A^T_{678}R^{678}+A^T_{91011}R^{91011}+A^T_{67891011}R^{67891011} \equiv \\
&\qquad((1-N_1){N_2}^{-\frac{1}{2}}\sin{\xi})R^{678}+i((1-N_2)N_1^{-\frac{1}{2}}\tan{\xi})R^{91011}\label{dyonicmembranegaugefields} \\
\nonumber &\qquad +((1-N_1)(N_1N_2)^{\frac{1}{2}}\cos{\xi})R^{6\ldots 11}
\end{align}
Where $T$ indicates that the gauge field is given in tangent space coordinates. Note that in this form when $\xi=0$ the interpolating group element reduces to the dual form of the group element for a single $M5$ brane.

To extract the field strength we premultiply each active gauge field component $A^T$ by the appropriate product of vielbeins ${e_{\mu}}^{a}$, which we read off from the volume element \ref{dyonicmembranemetric} to find the gauge field component in world volume indices. Explicitly,
\begin{align}
A^W_{678}&={e_{6}}^{6}{e_{7}}^{7}{e_{8}}^{8}A_{678}^T=(N^{-1}_1-1)\sin{\xi}\\
A^W_{91011}&={e_{9}}^{9}{e_{10}}^{10}{e_{11}}^{11}A_{91011}^T=i(N^{-1}_2-1)\tan{\xi} \\
A^W_{6\ldots 11}&={e_{6}}^{6}\ldots{e_{11}}^{11}A_{6\ldots 11}^T=(1-N_1)\cos{\xi}
\end{align}
These potentials are the components of two three-forms and a six-form. The six index gauge potential $A^W_{6\ldots 11}$ sources a dual four-form field strength. To find the field strength we take the exterior derivative of the relevant forms to find:
\begin{align}
\nonumber {\cal F}_{4(M2\oplus M5)}=&d(N^{-1}_1)\sin{\xi} \wedge dx_6\wedge dx_7\wedge dx_8-\star d(N_1)\cos{\xi}\\
&-\frac{i}{2N_2^2}d(N_1)\sin{2\xi} \wedge dx_9\wedge dx_{10}\wedge dx_{11}
\end{align}
The exterior derivative and the Hodge dual ($\star$) act in the five-dimensional subspace with coordinates $\{x_1,\ldots x_5\}$,
which is mutually transverse to the world volumes of the M2 and S2 solution worldvolumes. 
The solution may be described as a bound state of an $M2(678)$ and an $S2(91011)$ brane, or in the more usual language as the bound state of an $M2(678)$ and an $M5(67891011)$ brane. Due to the non-zero commutator of $[R^{678},R^{91011}]$ we find that the field strength is no longer a simple sum of the field strengths for the constituent M2 and S2 branes, instead we find additional terms in the field strength that give rise to a binding energy. 
The harmonic functions of the solution are derived from the harmonic function of the membrane smeared in the directions longitudinal to the constituent brane worldvolumes, which is the prescription for harmonic superposition \cite{Tseytlin:1996p1469}. This was implicit in the ansatz \ref{ratio} where $N_1$ and $N_2$ are harmonic functions of $\{x_1,x_2,x_3,x_4,x_5\}$.

The dyonic membrane illustrates and guides our principle observation, that one may extract marginal bound state solutions from the roots of the adjoint algebra of $E_{11}$. Let us state the basic steps:
\begin{list}{$\bullet$}{}
\item{Roots are decomposed into sums of roots associated to marginal solutions; in the limit into sums of M2 and S2 brane roots.}
\item{Each solution is smeared along the longitudinal directions of \emph{all} branes, allowing all harmonic functions to be related to one of the harmonic functions by angle parameters. 
\item{The constituent branes and superposed.}
\item{The gauge field is determined by the conditions that: \\  
- in the limits of the angle parameters it reduces to lower level marginal or non-marginal solutions, and in particular to individual brane solutions; \\
- interchange of constituent gauge fields and the corresponding generators, which do not effect the commutators required to obtain the solution, leave the overall gauge field unaltered;\\
- upon conversion to world volume indices using the veilbein of the background of the bound state, the gauge fields may be written using the harmonic function and angle parameters which interpolate between the constituent electric brane states. Magnetic S-brane states will have imaginary gauge fields and a singular limit in the angle parameters.}}
\end{list}
Using this set-up we can find other non-marginal states from the $E_{11}$ algebra.
\end{subsection}
\begin{subsection}{The transversely boosted M2 and M5 branes}
Amongst the marginal M-brane solutions are the longitudinally boosted M2 and M5 branes, but amongst the non-marginal solutions we can find the transversely boosted branes \cite{Russo:1996p693}. These branes are constructed from a KK-wave root $\beta_{pp}$ such that $\beta_{pp}\cdot\beta_{M2}=\beta_{pp}\cdot\beta_{M2}=-1$. 

Consider the following root, corresponding to a boosted M2 brane:
\begin{equation}
\beta_{M2+}=e_8+e_9+e_{10}=\beta_{M2}+\beta_{pp}\qquad \mbox{where}\quad \beta_{KK}=e_8-e_{11}
\end{equation}
Following the prescription we smear the KK-wave along the M2 brane directions so that the harmonic functions of the solution are:
\begin{equation}
N_1=1+\frac{Q}{r^6},\qquad N_2=1+\frac{Q\cos^2{\xi}}{r^6}\qquad \mbox{where}\quad r^2=x_1^2+\ldots x_7^2+x_{11}^2
\end{equation}
The composite roots $\beta_{pp}$ and $\beta_{M2}$ give the diagonal components of the metric:
\begin{align}
d\hat{s}_{pp}^2&=-N_1^{-1}dt_8^2+N_1dx_{11}^2+d\Omega_9^2\\
\nonumber ds_{S2}^2&=N_2^{\frac{1}{3}}(N_2^{-1}(dx_9^2+dx_{10}^2+dx_{11}^2)+d\Omega_{(1,7)}^2) 
\end{align}
Where we indicate by $d\hat{s}$ that this is not the full line element for the KK-wave only the diagonal parts of the metric are indicated. 
The superposed diagonal solutions give the crucial vielbein components that will be used shortly to reconstruct the off-diagonal part of the metric from the group element in this case:
\begin{align}
d\hat{s}_{M2+}^2&=N_2^{\frac{1}{3}}(N_2^{-1}(dx_9^2+dx_{10}^2+N_1dx_{11}^2)-N_1^{-1}dt_8^2+dy_idy^i) \label{diagonalboostedM2}
\end{align}
To find the group element (\ref{interpolatinggroupelement}) we calculate the following inner products:
\begin{equation}
\beta_{KK}\cdot \beta_{M2_1}=-1 \qquad \beta_{KK}\cdot \beta_{M2_2}=1 \qquad \beta_{M2_1}\cdot \beta_{M2_2}=1
\end{equation}
The group element for this solution has the particular form:
\begin{align}
\nonumber g_{\beta}=&\exp{(-\frac{1}{\beta^2}\ln{N_1}(H\cdot \beta_{pp})-\frac{1}{\beta^2}\ln{N_2}(H\cdot \beta_{M2}))}\\
&\exp((1-N_1)^{\frac{1}{2}}(1-\frac{N_1}{N_2})^{\frac{1}{2}}{K^{8}}_{11}+(1-N_2)^{\frac{1}{2}}(1-\frac{N_2}{N_1})^{\frac{1}{2}}R^{91011}\label{boostedM2groupelement} \\
\nonumber &\qquad +(1-N_1)^{\frac{1}{2}}(1-N_2)^\frac{1}{2}(\frac{N_1}{N_2})^{\frac{1}{2}}R^{8910})
\end{align}
Using the veilbein of (\ref{diagonalboostedM2}) to write the gauge components in terms of worldvolume coordinates we find an off-diagonal component of the metric and a three-form gauge field respectively:
\begin{align}
{h^{11}}_8^{W}&=(N_1^{-1}-1)\sin{\xi} \label{offdiagonal}\\
\nonumber A^{W}_3&=(N_2^{-1}-1)\tan{\xi}dx_9\wedge dx_{10}\wedge dx_{11}\\
&\qquad +(1-N_1)N_2^{-1}\cos{\xi}dt_8\wedge dx_9\wedge dx_{10} \label{threeform}
\end{align}
Finally including the off-diagonal metric component (\ref{offdiagonal}) gives the volume element of the transversely boosted $M2$ brane \cite{Russo:1996p693}:
\begin{align}
\nonumber ds_{M2+}^2&=N_2^{\frac{1}{3}}(N_2^{-1}(dx_9^2+dx_{10}^2+N_1(dx_{11}-(N_1^{-1}-1)\sin{\xi}dt_8)^2)-N_1^{-1}dt_8^2\\
&\qquad +dy_idy^i) \label{boostedM2}
\end{align}
By writing $N_1=1+W$ and $\xi=\theta+\frac{\pi}{2}$ we find the transversely boosted $M2$ brane in the variables of \cite{Russo:1996p693}. We recall the observation here of \cite{Russo:1996p693} that this solution may be written more simply, and more obviously, as a boosted membrane by a Lorentz transformation of the coordinates, where the boost parameter is $\xi$. The solution interpolates between the KK-wave ($\xi=\frac{\pi}{2}$) and an M2-brane solution ($\xi=0$) but $\xi$ now has the physical interpretation of a boost parameter with the Lorentz transformation given by:
\begin{equation}
x_{11}\to \frac{1}{\cos{\xi}}(\bar{x}_{11}-\sin{\xi}\bar{t}_8), \quad t_{8}\to \frac{1}{\cos{\xi}}(\bar{t}_{8}-\sin{\xi}\bar{x}_{11})
\end{equation}
The transversely boosted $M5$ solution is recovered in a similar way from a group element analogous to \ref{boostedM2groupelement}.
\comment{
\begin{align}
\nonumber g_{\beta}=&\exp{(-\frac{1}{\beta^2}\ln{N_1}(H\cdot \beta_{pp})-\frac{1}{\beta^2}\ln{N_2}(H\cdot \beta_{M5}))}\\
&\exp((1-N_1)^{\frac{1}{2}}(1-\frac{N_1}{N_2})^{\frac{1}{2}}{K^{5}}_{11}+(1-N_2)^{\frac{1}{2}}(1-\frac{N_2}{N_1})^{\frac{1}{2}}R^{67891011}\label{boostedM5groupelement} \\
\nonumber &\qquad -(1-N_1)^{\frac{1}{2}}(1-N_2)^\frac{1}{2}(\frac{N_1}{N_2})^{\frac{1}{2}}R^{5678910})
\end{align}
Where,
\begin{align}
\beta_{pp}=e_5-e_{11} &\qquad \beta_{M5}=e_6+\ldots+e_{11}\\
N_1=1+\frac{P}{r^3} &\qquad N_2=1+\frac{P\cos{\xi}^2}{r^3}\\ 
r^2=x_1^2&+x_2^2+\ldots x_4^2+x_{11}^2
\end{align}
The volume element is:
\begin{align}
\nonumber ds_{M5+}^2&=N_2^{\frac{2}{3}}(N_2^{-1}(dx_6^2+\ldots+dx_{10}^2+N_1(dx_{11}-(N_1^{-1}-1)\sin{\xi})^2)\\
&\qquad \qquad-N_1^{-1}dt_5^2+dy_idy^i) \label{boostedM5}
\end{align}}
\end{subsection}
\end{section}

\begin{section}{The $E_{11}$ guide to Exotic Solutions}
\begin{subsection}{The KK-branes of M-theory}
In this section we derive the M-theory KK-brane solutions of  \cite{LozanoTellechea:2001p1595} from the associated roots of the adjoint representation of $E_{11}$. The existence of two KK-branes in M-theory, the $M2_6$ and the $M5_3$, has been argued for by oxidising ten-dimensional KK-branes of ten-dimensional string theory (which were derived using $S$-duality) \cite{LozanoTellechea:2001p1595}. They also arise from the U-duality transformations of M-theory \cite{Cook:2008p936} which correspond to Weyl reflections of $E_{11}$. 

A useful tool for identifying the roots of $E_{11}$ associated to the KK-branes of M-theory is the mass formula\footnote{A similar mass formula for $E_{10}$ was used in \cite{Brown:2004p2806}.} identified in \cite{Cook:2008p936} from which the masses of the $M2_6$ and the $M5_3$ may be associated to particular roots appearing in the $l_1$, or charge, representation of $E_{11}$. This formula, after division by the brane volume, correctly reproduced all the brane tensions of eleven dimensional supergravity, as well as the tensions of the Dp-branes of the IIA and IIB string theories together with the correct powers of the string coupling constant, $g_s$, and $\alpha'$. Furthermore it gave agreement with the tensions of exotic objects which had been predicted as a consequence of the U-duality symmetry applied to known brane charges \cite{Elitzur:1998p319,Obers:1998p275,Obers:1999p939}. The mass of the exotic objects is a monomial function of the radii of compactification with at least one of the radii appearing non-linearly - we will use this as our definition of a KK-brane. For example, the mass of the KK6 brane, which can be seen in the particle/flux multiplet of \cite{Elitzur:1998p319,Obers:1998p275,Obers:1999p939} and is,
\begin{equation}
{\cal M}_{KK6}=\frac{R_4R_5R_6R_7R_8R_9R_{10}R_{11}^2}{l_p^9}
\end{equation}
The radius $R_{11}$ is squared in this mass, so we define it to be a KK-brane. The mass formula of  \cite{Cook:2008p936} was given as a map from the charge algebra, or  $l_1$ representation, of $E_{11}$. Previously \cite{Kleinschmidt:2004p62} an injective map was constructed which took roots of the adjoint of $E_{11}$, whose generators are associated to gauge fields in the physical theory, into the weights of the $l_1$ representation, whose generators are associated to brane charges in the physical theory. This injective map removed a single index from the generators of the $E_{11}$ adjoint in a systematic fashion. Thus, the three form $R^{a_1a_2a_3}$ was mapped to a two-form charge $Z^{a_1a_2}$, the six-form generator to a five-form charge $Z^{a_1\ldots a_5}$ and so on. It is therefore a straightforward step to give a mass formula that acts directly upon the adjoint of $E_{11}$\footnote{However we realise that while this is a useful shortcut we should not lose track of the physical notion that the mass is derived from the the charge algebra, the $l_1$ representation.}. Given a root, $\beta$, of $E_{11}$ expressed in the ${e_i}$ basis:
\begin{equation}
\beta=\sum_{i=1}^{11}b_ie_i
\end{equation}
The mass formula in the compact setting is:
\begin{equation}
{\cal M'}_\beta=\Pi_{i=1}^{11}(\frac{R_i}{l_p})^{b_i}
\end{equation} 
Where $R_i$ are the radii of the compact directions. The prime is used here to indicate that all longitudinal directions have been compactified. When a direction, $x^i$, is not compactified the constant $R_i$ is set to one  \cite{Cook:2008p936}. For example, the $M2$ brane is associated to the root $\alpha_{11}=e_9+e_{10}+e_{11}$ and its mass derived from the root is,
\begin{equation}
{\cal M'}_{M2}=\frac{R_9R_{10}R_{11}}{l_p^3}
\end{equation}
However at least one of the longitudinal directions of the M2 brane is timelike and it is natural not to compactify this direction, if we associate the direction $x^{11}$ to time and decompactify, then we set $R_{11}=1$ to obtain,
\begin{equation}
{\cal M}_{M2}=\frac{R_9R_{10}}{l_p^3}
\end{equation}
In general we will not consider the compactification of the timelike direction, so let us rewrite our mass formula to take into account a non-compact timelike direction,
\begin{equation}
{\cal M}_\beta=(\frac{1}{l_p})^{b_t}\Pi_{i=1}^{10}(\frac{R_i}{l_p})^{b_i}
\end{equation} 
Where we have used $b_t$ to indicate the coefficient of the timelike coordinate of $e_t$ in the root. The $E_{11}$ adjoint algebra is expressed in terms of representations of $SL(11)$ by the deletion of the root $\alpha_{11}$. The number of times $\alpha_{11}$ must be deleted from a root in $E_{11}$ to arrive at a highest weight of $SL(11)$ is called the level and denoted $m_{11}$ here. As a consequence of the formulae of  \cite{Cook:2008p936} one can express the mass formula in terms of the level as,
\begin{equation}
{\cal M}_\beta=\frac{\Pi_{i=1}^{10}{R_i}^{b_i}}{l_p^{3m_{11}}} \label{masseq}
\end{equation} 
This formula is very useful, for example we can use it to immediately locate the root associated to the $M2_6$ solution and the $M5_3$ solution, by searching for their masses, which are \cite{LozanoTellechea:2001p1595},
\begin{align}
{\cal M}_{M2_6}&=\frac{R_4R_5(R_6\ldots R_{11})^2}{l_p^{15}}\label{M26mass} \\
{\cal M}_{M5_3}&=\frac{R_4\ldots R_8(R_9\ldots R_{11})^2}{l_p^{12}}\label{M53mass}
\end{align}
Referring to equation (\ref{masseq}), we find the corresponding roots, both appearing in the adjoint of $E_{11}$, are,
\begin{align}
\beta_{M2_6}&=(0,0,1,1,1,2,2,2,2,2,2)_{e_i}=(0,0,1,2,3,5,7,9,6,3,5)_{\alpha_i}\label{M26root} \\
\beta_{M5_3}&=(0,0,1,1,1,1,1,1,2,2,2)_{e_i}=(0,0,1,2,3,4,5,6,4,2,4)_{\alpha_i}\label{M53root}
\end{align}
Where we have expressed the roots in both the $e_i$ coordinate basis and the $\alpha_i$ simple root basis.  For reference we also draw the Young tableaux of the associated generators:
\begin{align}
\Yvcentermath1 
&\yng(2,2,2,2,2,2,1,1,1) \hspace{60pt} &\yng(2,2,2,1,1,1,1,1,1)\label{KKyoungtableaux}\\
\nonumber &M2_6 &M5_3
\end{align}
There are a number of tables of adjoint roots of $E_{11}$ in the literature, the first appearing in \cite{Nicolai:2002p1596}, where one can verify that the outer multiplicity of these roots is not zero. However for our purposes it is most useful to see the roots expressed in the $e_i$ basis, and for this reason the reader is also referred to the tables of \cite{Cook:2007p642}, to confirm these roots do appear in the root tables of the adjoint of $E_{11}$. Alternatively the reader may prefer to use the simple methods outlined in section one.

These roots may be expressed as sums of lower level roots in a number of ways. In both the case of the $M2_6$ and the $M5_3$ the root may be written as a sum of roots $(\beta_{pp}, \beta_{M2}, \beta_{M5})$. Our aim is to construct the Young tableaux of equation (\ref{KKyoungtableaux}) as a combination of the Young tableaux of the M2 and the M5 brane,
\begin{equation}
\Yvcentermath1 
\yng(1,1,1) \qquad \text{and} \qquad \yng(1,1,1,1,1,1)\quad,
\end{equation}
together with coordinate rotations corresponding to roots in the local sub-algebra.
\subsubsection{The $M2_6$ brane}
Let us begin with the $M2_6$ solution. We observe that,
\begin{equation}
\beta_{M2_6}=\beta_{M2}+2\beta_{M5}+\beta_{pp*}
\end{equation}
Where $\beta_{pp*}$ indicates a sum of roots from the local sub-algebra and we deduce that,
\begin{equation}
\beta_{pp*}= (0,0,1,1,1,0,0,0,-1,-1,-1)_{e_i}=(0,0,1,2,3,3,3,3,2,1,0)_{\alpha_i} \label{pp*}
\end{equation}
The root for the $M2_6$ brane is comprised of two copies of the root for the M5 brane and one copy of the root associated to the M2 brane together with the application of the local generators associated to $\beta_{pp*}$ which rotate the coordinates $\{x^9,x^{10},x^{11}\}$ to $\{x^3,x^4,x^5\}$. We recall that the action of the local sub-algebra is to rotate the time coordinate \cite{Keurentjes:2004p1580}, meaning in this case that this composite solution may be interpreted as either $M2\oplus S5 \oplus S5$ or $M5\oplus S2\oplus S5$.
In terms of representations of $A_{10}$ the $M2_6$ corresponds to the irreducible representation with highest weight $\lambda_2+\lambda_8$, by the identification above we are stating that this representation occurs within the representation whose highest weight is $\lambda_5+2\lambda_8$. We will consider the $M2\oplus S5 \oplus S5$ interpretation here.

Following the procedure described in section 3 we smear the constituent brane solutions over their relative transverse directions to obtain three harmonic functions. However as the transverse space is two-dimensional it is impossible to find three real harmonic functions such that $1-M_1M_2=1-N$, instead we use the following harmonic, holomorphic and anti-holomorphic functions:
\begin{align}
\nonumber N&=1+Q\ln{r}, \\
M_1&=1+Q\ln{\omega}\cos^2{\xi}, \quad M_2=1+Q\ln{\bar{\omega}}\cos^2{\xi} \label{M26functions}\\
\nonumber & \Rightarrow \qquad \cos{\xi}=\sqrt{\frac{1-\frac{1}{2}(M_1+M_2)}{1-N}}
\end{align}
Where $r^2=x_1^2+x_2^2=\omega\bar{\omega}$, with $\omega={x_1+ix_2}$. The solution interpolates between the membrane ($\xi=\frac{\pi}{2}$) and the exotic $M2_6$ ($\xi=0$) brane. 
The function $N$ is harmonic in the overall two-dimensional transverse space, but the functions $M_1$ and $M_2$ are holomorphic and antiholomorphic respectively \cite{LozanoTellechea:2001p1595}. The $M2_6$ solution has also been derived from a Kac-Moody symmetry following a different method in \cite{Englert:2007p605}. The metric for the full solution is constructed from just the diagonal elements of the metric and may now be read off using the roots for the constituent solutions, $\beta_{M2}$ and $\beta_{S5}$:
\begin{align}
\beta_{M2}=e_3+e_4+e_5, \qquad \beta_{S5}=e_6+\ldots e_{11}
\end{align}
Superposing the three component solutions we obtain:
\begin{align}
\nonumber ds_{M2_6}^2&={N}^{\frac{1}{3}}{(M_1M_2)}^{\frac{2}{3}}(dx_1^2+dx_2^2+N^{-1}(-{dt_3}^2+{dx_{4}}^2+{dx_{5}}^2)\\
&\qquad +(M_1M_2)^{-1}(dx_6^2+\ldots +dx_{11}^2)) \label{M262}
\end{align}
The group element for this solution is short, principally because $[R^{345},R^{6\ldots11}]=0$ and $[R^{6\ldots 11},R^{6\ldots 11}]$=0. This solution is in a different class to those we have considered hitherto. In the earlier examples we considered sets of generators whose commutators gave the generator of interest. More precisely if we had followed the same procedure here we would have considered three different generators such as $R^{6\ldots 11}$, $R^{5611}$ and $R^{3910}$ whose commutators generate $R^{3\ldots 11,91011}$. The group element would have been parameterised by three variables: $r$ and two angle variables. With this example we see that it is possible to find alternative group elements whose form differs from that of equation (\ref{interpolatinggroupelement}) but recreates M-brane solutions. We propose the group element: 
\begin{align}
g_{\beta}=&\exp{(-\frac{1}{\beta^2}\ln{N}(H\cdot \beta_{M2})-\frac{1}{\beta^2}\ln{(M_1M_2)}(H\cdot \beta_{S5}))}\label{M26}\\
\nonumber &\exp((1-\frac{N}{M_1M_2})R^{345}+(1-\frac{\sqrt{M_1M_2}}{N})^\frac{1}{2}(1+\frac{\sqrt{M_1M_2}}{N})^\frac{1}{2}R^{67891011})
\end{align}
Once the tangent space indices of the gauge fields are transformed into wordlvolume indices we find:
\begin{align}
A^W_{345}=\frac{M_1M_2}{N}-1 \qquad A^W_{6\ldots 11}=\frac{-B}{M_1M_2}
\end{align}
Where we use $B\equiv \frac{1}{2}(M_1-M_2)$ and the following identity:
\begin{align}
M_1M_2&=(1-(1-N)\cos^2{\xi})^2+(B\cos^2{\xi})^2\\
&=\left(\frac{M_1+{M}_2}{2}\right)^2+\left(\frac{M_1-{M}_2}{2}\right)^2
\end{align}
The solution interpolates between the $M2$ and the $M2_6$ brane. In reproducing the line element of the $M2_6$ from the $E_{11}$ fields we have gained an understanding of the $M2_6$ as a composite solution. The line element of the $M2_6$ consists of two S5 branes and an M2 solution. One could also consider the Wick rotation of the solution which would consist of two M5 branes and an S2 brane, or indeed attempt to understand this solution in terms of the group element of (\ref{interpolatinggroupelement}). 

\subsubsection{The $M5_3$ brane}
Let us repeat the process of finding the line element for the $M5_3$ KK-brane from the $E_{11}$ group element. Using the candidate root of $E_{11}$ associated to the $M5_3$ given in equation \ref{M53root}. Proceeding as before we note that:
\begin{equation}
\beta_{M5_3}=2\beta_{M2}+\beta_{M5}+\beta_{pp*}
\end{equation}
Proceeding as before we associate the coordinate rotation associated to $\beta_{pp*}$ to the gauge fields of the M5 brane or to one of the gauge fields of the pair of M2 branes, giving $M5\oplus S2 \oplus S2$ and $M2\oplus S5 \oplus S5$ respectively. As before we focus on the former case and identify it with the $M5_3$ brane of \cite{LozanoTellechea:2001p1595}. By writing the individual metrics and harmonically superposing the solutions we arrive at the line element:
\begin{align}
\nonumber ds_{M5_3}^2&={M}^{\frac{2}{3}}{(N_1N_2)}^{\frac{1}{3}}(dx_1^2+dx_2^2) +{(\frac{M}{N_1N_2})}^{-\frac{1}{3}}(-dt_{3}^2+dx_4^2+\ldots +dx_{8}^2)\\
&\qquad +{(\frac{M}{N_1N_2})}^{\frac{2}{3}}({dx_9}^2+\ldots +{dx_{11}}^2)\label{M531}
\end{align}
The $M5_3$ solution is reproduced from the group element analogous to that used for the $M2_6$ solution \ref{M26}. The functions $M$, $N_1$ and $N_2$ are identified as in \ref{M26functions} under an interchange of the labels $N$ and $M$.

\subsubsection{An exotic pure gravity solution: $WM_7$}
One other exotic M-theory solution is derived in \cite{LozanoTellechea:2001p1595} in addition to the $M2_6$ and $M5_3$ branes, denoted the $WM_7$ which is a deformation of the $KK$ wave. Its mass is:
\begin{equation}
{\cal M}_{WM_7}=\frac{(R_4\ldots R_{10})^2R_{11}^3}{l_p^{18}}
\end{equation}
Hence it is expected to be derived from the level 6 root:
\begin{equation}
\beta_{WM_7}=e_3+2(e_4+\ldots+ e_{10})+3e_{11}=(0,0,1,3,5,7,9,11,7,3,6)_{\alpha_i}
\end{equation}
This root is associated to the highest weight Young tableau:
\begin{equation}
\Yvcentermath1 
\yng(3,2,2,2,2,2,2,2,1)
\end{equation}
We observe that:
\begin{equation}
\beta_{WM_7}=2\beta_{KK6}+\beta_{KK} \qquad  \mbox{where } \quad \beta_{KK}=e_3-e_{11}
\end{equation}
For this solution we may use a short-cut via Hodge duality to find the solution hence the group element encoding the solution is:
\begin{align}
g_{\beta}=&\exp{(-\frac{1}{\beta^2}\ln{N}(H\cdot \beta_{KK})-\frac{1}{\beta^2}\ln{(M_1M_2)}(H\cdot \beta_{\beta_{KK6}}))}\label{WM7}\\
\nonumber &\exp((1-\frac{N}{M_1M_2}){K^3}_{11})
\end{align}
Following our method the functions $M_1$, $M_2$ and $N$ are smeared so that they are harmonic in the directions $x_1$ and $x_2$, as in the previous examples. The functions are identical to those given in equation (\ref{M26functions}). In this example we demonstrate how it is possible to take advantage of Hodge dualisation to remove antisymmetric blocks of nine boxes from the Young tableau. The generator has been "dualised" twice using the method outlined in section 3, that is we have used the observation that:
\begin{align}
2\beta_{9}-\beta_{WM_7}=\beta_{KK}
\end{align}
The volume element encoded in the group element (\ref{WM7}) is:
\begin{align}
ds_{WM_7}^2&=(M_1M_2)(dx_1^2+dx_2^2)-\left(\frac{N}{M_1M_2}\right)^{-1}dt_3^2\\
\nonumber &+\frac{N}{M_1M_2}(dx_{11}-(\frac{M_1M_2}{N}-1)dt_3)^2+d\Omega_{7}^2
\end{align}
Recalling that,
\begin{equation}
M_1M_2=(1-(1-N)\cos^2{\xi})^2+(B\cos^2{\xi})^2
\end{equation}
We see that it interpolates between the KK wave and the $WM_7$ brane. There is a tower of similar roots in the $E_{11}$ root system, the so-called dual roots \cite{Riccioni:2006p600}, for which we can readily write down a group element and metric. Consider the set of dual roots made from dressing the $KK6$ root up with multiple copies of $\beta_9$, i.e.
\begin{equation}
\beta_{KK6^n}=\beta_{KK6}+n\beta_9
\end{equation}
Such that the root is dualised to:
\begin{equation}
n\beta_9-\beta_{KK6^n}=\beta_{KK}
\end{equation}
We predict a group element:
\begin{align}
g_{\beta}=&\exp{(-\frac{1}{\beta^2}\ln{N}(H\cdot \beta_{KK})-\frac{1}{\beta^2}\ln{(M_1\ldots M_{n+1})}(H\cdot \beta_{\beta_{KK6}}))}\label{KK6n}\\
\nonumber &\exp((1-\frac{N}{M_1\ldots M_{n+1}}){K^3}_{11})
\end{align}
That encodes the metric:
\begin{align}
ds_{KK6^n}^2&=(M_1\ldots M_{n+1})(dx_1^2+dx_2^2)-\left(\frac{N}{M_1\ldots M_{n+1}}\right)^{-1}dt_3^2\\
\nonumber &+\frac{N}{M_1\ldots M_{n+1}}(dx_{11}-(\frac{M_1\ldots M_{n+1}}{N}-1)dt_3)^2+d\Omega_{7}^2
\end{align}
The metric corresponds to a deformation of the KK6 brane. However the exact format of the harmonic functions and the interpolating angles remains to be expressed.
\end{subsection}

\begin{subsection}{The KK6 brane revisited}
In the previous sections we have investigated a solution generating technique using simple group elements to reproduce known composite solutions from higher level roots associated to the Borel sub-algebra of the adjoint representation of $E_{11}$. Through this method we have given a consistent interpretation of exotic branes as bound states of the fundamental M-branes. However the method of partitioning a root into sums of lower level roots, which are in turn associated to brane solutions, may give more than one interpretation of the consituent states. To highlight this point we examine again the level three root associated to the KK6 brane. In the first instance we will interpret it as a bound states of two M-branes before suggesting a group element consisting of three membrane charges and two interpolation angles. 

The level four root $\beta_{KK6}=e_4+\ldots e_{10}+2e_{11}$, which we saw earlier was associated to the KK6 monopole and the embedded Taub-NUT solution is associated to the generator $R^{456\ldots 11,11}$ and may be partitioned into roots from lower levels in two different ways:
\begin{align}
\beta_{KK6}=&3\beta_{M2}+\beta^{(1)}_{pp} \label{level3b}\\
=&\beta_{M5}+\beta_{M2}+\beta^{(2)}_{pp} \label{level3c}
\end{align}
Where,
\begin{align}
\beta^{(1)}_{pp}=&e_4+e_5+e_6+e_7+e_8-2e_9-2e_{10}-e_{11}\\
\nonumber \beta^{(2)}_{pp}=&e_4+e_5-e_9-e_{10}
\end{align}
The roots $\beta^{(i)}_{pp}$ correspond to multiple applications of $SL(11)$ generators ${K^a}_b$ and have the effect of rotating the active coordinates of the gauge field. For example $\beta^{(2)}_{pp}$ rotates the coordinates $\{x_9,x_{10}\}$ into $\{x_4,x_5\}$ and $\beta^{(1)}_{pp}$ rotates the coordinates $\{x_9,x_9,x_{10},x_{10},x_{11}\}$ into $\{x_4,x_5,x_6,x_7,x_8\}$. 

Consider the partition of $\beta_{KK6}$ given in (\ref{level3c}), the rotation associated to $\beta^{(2)}_{pp}$ could be applied to either the $M5$ solution or the $M2$ solution. The rotation has the effect of transforming one of the brane solutions while ensuring it remains an electric solution \cite{Keurentjes:2004p1580}, while the unrotated solutions will be spacelike, or magnetic, versions of the brane solutions. We have two choices from (\ref{level3c}):
\begin{align}
\beta_{KK6}=\beta_{M5}+\beta_{S2}=\beta_{S5}+\beta_{M2}
\end{align}
Let us consider the partition into an M5 and an S2 root. Taking the M5 to be aligned along $\{t_4,x_5,\ldots x_{8},x_{11}\}$ and the S2 along $\{x_9,x_{10},x_{11}\}$ the superposed volume element is:
\begin{align}
\nonumber ds_{S2\oplus M5}^2=&M^{\frac{2}{3}}N^{\frac{1}{3}}(dx_1^2+\ldots dx_3^2+M^{-1}(-dt_4^2+\ldots+dx_{8}^2)\\
&+N^{-1}(dx_{9}^2+dx_{10}^2)+M^{-1}N^{-1}dx_{11}^2)
\end{align}
We can interpret the solution as an $M5$-brane intersecting an $S2$-brane at a point.
Noting that when $N=1$ the solution is that of an electric M5-brane, while when $M=1$ the solution is that of a magnetic spacelike S2-brane solution, as expected by our construction. Furthermore when $N=M$ the solution becomes that of equation (\ref{KK6dualgroupelement}). The interpolating, two-parameter group element \ref{interpolatinggroupelement} for these roots is:
\begin{align}
\nonumber g_{\beta}=&\exp{(-\frac{1}{\beta^2}\ln{M}(H\cdot \beta_{M5})-\frac{1}{\beta^2}\ln{N}(H\cdot \beta_{S2}))}\\
&\exp((1-M)^{\frac{1}{2}}(1-\frac{M}{N})^{\frac{1}{2}}R^{4567811}+(1-N)^{\frac{1}{2}}(1-\frac{N}{M})^{\frac{1}{2}}R^{91011} \\
\nonumber &\qquad +(1-M)^{\frac{1}{2}}(1-N)^{\frac{1}{2}}(\frac{M}{N})^{\frac{1}{2}}R^{4\ldots 11,11})
\end{align}
As before we have introduced an angle variable relating the ratio of the charges of the solutions:
$$\frac{1-N}{1-M}=\cos^2{\xi}$$
The field strength for this interpolating solution according to the two-parameter group element ansatz is:
\begin{align}
{\cal F}_4=&(\sin{\xi}) \star dM^{-1}\wedge dt_4\wedge \ldots \wedge dx_8 \wedge dx_{11}\\
\nonumber & -\frac{i}{2N^2}(\sin{2\xi})dM\wedge dx_9\wedge dx_{10} \wedge dx_{11} 
\end{align}
We may write the metric for this solution in terms of one harmonic function, $M$, which interpolates between an $M5$ brane background and the pure gravitational solution of the $KK6$ as:
\begin{align}
\nonumber ds_{M5\to KK6}^2=&(M^2(\sin^2{\xi}+M\cos^2{\xi}))^{\frac{1}{3}}(dx_1^2+\ldots dx_3^2+M^{-1}(-dt_4^2+\ldots \\ 
&+dx_{8}^2)+(\sin^2{\xi}+M\cos^2{\xi})^{-1}((dx_{9}^2+dx_{10}^2)\\ 
\nonumber &+M^{-1}(dx_{11}+(1-M)\cos^2{\xi}dt_3)^2))
\end{align}
Given the interpretation of the dyonic membrane as a bound state of an $M2$ and an $M5$ brane, this presentation suggests that the $KK6$ may be understood as a bound state of two $M5$ branes intersecting along a four-brane.

Let us turn to the second partition of $\beta_{KK6}$ given in equation (\ref{level3b}), which, once we include the rotation due to $\beta^{(2)}_{pp}$, interprets the solution as an object composed of one $M2$ solution and two $S2$ solutions. The volume elements for these solutions are:
\begin{align}
\nonumber ds_{M2}^2=&N_1^{\frac{1}{3}}(dx_1^2+\ldots dx_3^2)+N_1^{-\frac{2}{3}}(-dt_4^2+\ldots+dx_{6}^2)+N_1^{\frac{1}{3}}(dx_7^2+\ldots dx_{11}^2)\\
ds_{S2_1}^2=&N_2^{\frac{1}{3}}(dx_1^2+\ldots dx_6^2+dx_9^2+dx_{10}^2)+N_2^{-\frac{2}{3}}(dx_7^2+dx_8^2+dx_{11}^2)\\
\nonumber ds_{S2_2}^2=&N_3^{\frac{1}{3}}(dx_1^2+\ldots dx_8^2)+N_3^{-\frac{2}{3}}(dx_9^2+dx_{10}^2+dx_{11}^2)
\end{align}
Superposing these solutions gives:
\begin{align}
\nonumber ds_{M2}^2=&(N_1N_2N_3)^{\frac{1}{3}}[(dx_1^2+\ldots dx_3^2)+N_1^{-1}(-dt_4^2+\ldots+dx_{6}^2)\\
&+N_2^{-1}(dx_7^2+dx_8^2)+N_3^{-1}(dx_9^2+dx_{10}^2)+(N_2N_3)^{-1}dx_{11}^2]
\end{align}
Which one can interpret as one membrane and two spacelike membranes which intersect over a point. When either $N_2=1$ or $N_3=1$ the solution becomes the dyonic membrane.

To encode this solution in a group element we will introduce two angle parameters, $\xi$ and$\upsilon$, defined by:
\begin{equation}
\frac{1-N_2}{1-N_1}=\cos^2{\xi}\qquad \qquad \frac{1-N_3}{1-N_1}=\cos^2{\upsilon}
\end{equation}
To derive the gauge part of the three-parameter solution-generating group element we are guided by the symmetry under interchange of the two component magnetic branes and by the constraint that upon $\upsilon \to \frac{\pi}{2}$ ($N_3\to 1$) or $\xi \to \frac{\pi}{2}$ ($N_2\to 1$) the solution is reduced to a dyonic membrane. One three parameter solution-generating group-element is generically:
\begin{align}
\nonumber g_{\beta}&=\exp{(-\frac{1}{\beta^2}\sum_{i=1}^3\ln{N_i}(H\cdot \beta_{M2_i}))}\\
\nonumber &\quad \exp(\sin{\xi}\sin{\nu}(1-N_1)N_2^{\frac{\beta_1\cdot \beta_2}{2}}N_3^{\frac{\beta_1\cdot \beta_3}{2}}E_{\beta_1}\\
\nonumber & \qquad +i\tan{\xi}\sin{\nu}(1-N_2)N_1^{\frac{\beta_2\cdot \beta_1}{2}}N_3^{\frac{\beta_2\cdot \beta_3}{2}}E_{\beta_2}\\
\nonumber &\qquad +i\tan{\nu}\sin{\xi}(1-N_3)N_1^{\frac{\beta_3\cdot \beta_1}{2}}N_2^{\frac{\beta_3\cdot \beta_2}{2}}E_{\beta_3}\\
\nonumber & \qquad +\cos{\xi}\sin{\nu}(1-N_1)N_1^{\frac{\beta_4\cdot \beta_1}{2}}N_2^{-\frac{\beta_4\cdot \beta_2}{2}}N_3^{\frac{\beta_4\cdot \beta_3}{2}}E_{\beta_4}\\
\nonumber & \qquad +\cos{\nu}\sin{\xi}(1-N_1)N_1^{\frac{\beta_5\cdot \beta_1}{2}}N_2^{\frac{\beta_5\cdot \beta_2}{2}}N_3^{-\frac{\beta_5\cdot \beta_3}{2}}E_{\beta_5}\\
& \qquad +\cos{\nu}\sin{\xi}(1-N_1)N_1^{\frac{\beta_6\cdot \beta_1}{2}}N_2^{\frac{\beta_6\cdot \beta_2}{2}}N_3^{-\frac{\beta_6\cdot \beta_3}{2}}E_{\beta_6})
\end{align}
Where $\beta_4=\beta_1+\beta_2$, $\beta_5=\beta_1+\beta_3$, $\beta_6=\beta_1+\beta_2+\beta_3$
For the particular example we consider here we find:
\begin{align}
\nonumber g_{\beta}&=\exp{(-\frac{1}{\beta^2}\sum_{i=1}^3\ln{N_i}(H\cdot \beta_{M2_i}))}\\
\nonumber &\quad \exp((1-\frac{N_1}{N_2})^{\frac{1}{2}}(1-\frac{N_1}{N_3})^{\frac{1}{2}}R^{456}\\
\nonumber & \qquad +(1-N_2)^{\frac{1}{2}}(1-\frac{N_2}{N_1})^{\frac{1}{2}}(1-\frac{1-N_3}{1-N_1})^{\frac{1}{2}}R^{7811}\\
\nonumber &\qquad +(1-N_3)^{\frac{1}{2}}(1-\frac{N_3}{N_1})^{\frac{1}{2}}(1-\frac{1-N_2}{1-N_1})^{\frac{1}{2}}R^{91011}\\
\nonumber & \qquad +(1-N_2)^{\frac{1}{2}}(1-\frac{N_1}{N_3})^\frac{1}{2}(\frac{N_1}{N_2})^\frac{1}{2}R^{4567811}\\
\nonumber & \qquad +(1-N_3)^{\frac{1}{2}}(1-\frac{N_1}{N_2})^\frac{1}{2}(\frac{N_1}{N_3})^\frac{1}{2}R^{45691011}\\
& \qquad +(1-N_2)^\frac{1}{2}(1-N_3)^\frac{1}{2}(\frac{N_2}{N_3})^\frac{1}{2}R^{4567891011,11})
\end{align}
The gauge field components when transformed into world volume indices are:
\begin{align}
\nonumber A^W_{456}&=(N_1^{-1}-1)\sin{\xi}\sin{\upsilon}\\
\nonumber A^W_{7811}&=i(N_2^{-1}-1)\tan{\xi}\sin{\upsilon}\\
\nonumber A^W_{91011}&=i(N_3^{-1}-1)\sin{\xi}\tan{\upsilon}\\
\nonumber A^W_{4567811}&=(1-N_1)N_2^{-1}\cos{\xi}\sin{\upsilon}\\
\nonumber A^W_{45691011}&=(1-N_1)N_3^{-1}\cos{\upsilon}\sin{\xi}\\
\nonumber A^W_{4567891011,11}&=(1-N_1)N_3^{-1}\cos{\xi}\cos{\upsilon}
\end{align}
\begin{figure}[h]
\centering
\includegraphics[scale=0.4,angle=0]{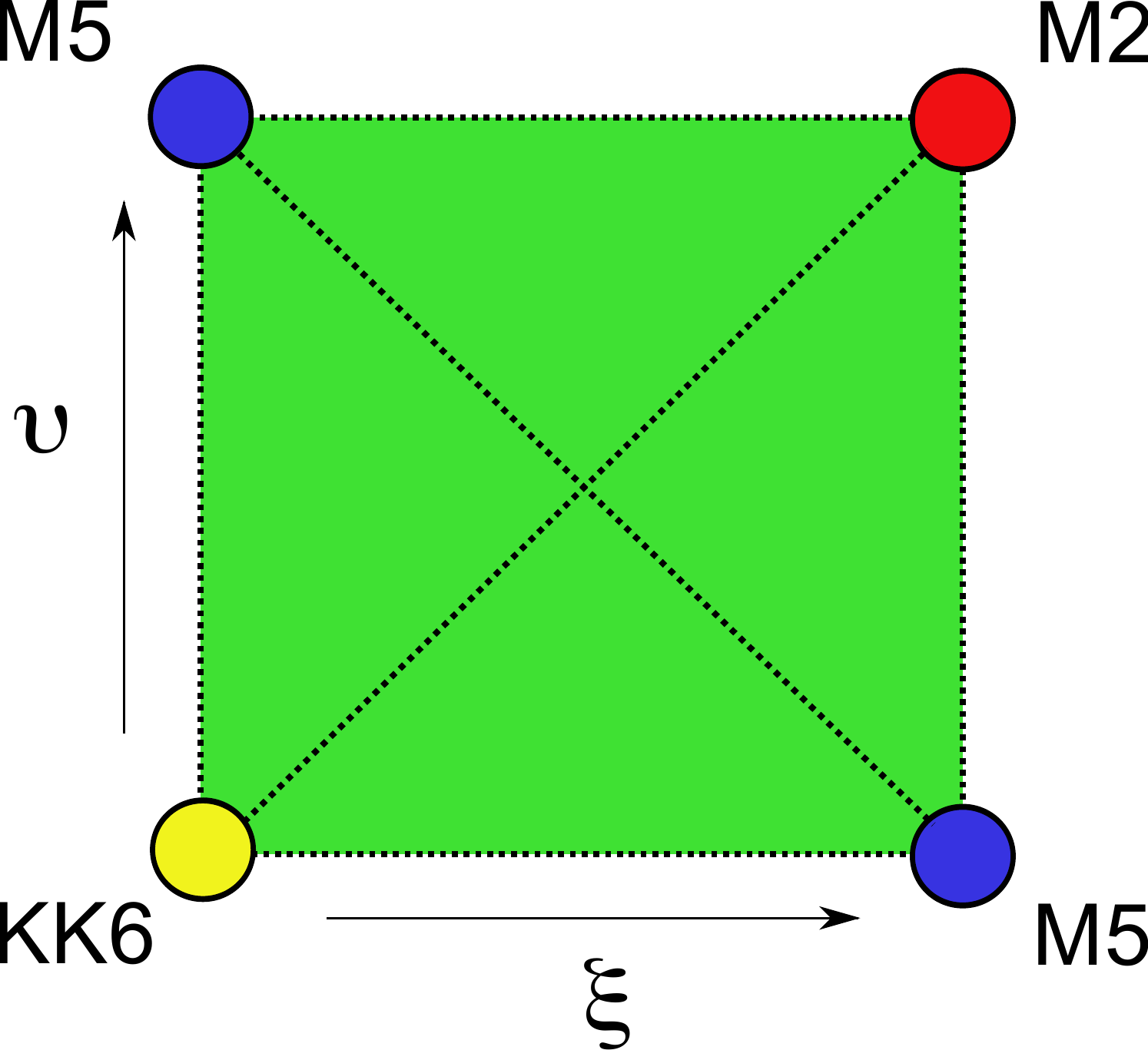}
\caption{\small{The decomposed KK6 solution. Along the horizontal $\xi$ varies from $0$ to $\frac{\pi}{2}$, while vertically $\upsilon$ varies from $0$ to $\frac{\pi}{2}$. Consequently the KK6 corner has $N_1=N_2=N_3$, the bottom right M5 brane has $N_2=1$ and $N_1=N_3$, the top left M5 brane has $N_3=1$ and $N_1=N_2$, the M2 brane in the top right corner has $N_2=N_3=1$.} \label{decomposedkk6}}
\end{figure}
The solution is invariant under the interchange of the harmonic functions associated to the S2 branes, $N_2$ and $N_3$, and the gauge field components are permuted accordingly. The ansatz for writing down the field strength requires us to recognise a priori which components are dualised and which are not. In the list above the first three fields are taken as components of a three form whose exterior derivative contributes to the four form field strength. The remaining fields are taken as dualised components: for the two six-forms the Hodge dual is taken in the five coordinates transverse to the field, however for the mixed-symmetry tensor the Hodge dual in the transverse coordinates gives a symmetric rank two tensor which contributes to the gravity sector and modifies the volume element of the solution. We find a four-form field strength and an off-diagonal vielbein component:
\begin{align}
\nonumber F_{4}=&d(N^{-1}_1)\sin{\xi}\sin{\upsilon} \wedge dt_4\wedge dx_5\wedge dx_6\\
&-\frac{i}{2N_2^2}d(N_1)\sin{2\xi}\sin{\upsilon} \wedge dx_7\wedge dx_{8}\wedge dx_{11}\\
\nonumber &-\frac{i}{2N_3^2}d(N_1)\sin{2\upsilon}\sin{\xi} \wedge dx_9\wedge dx_{10}\wedge dx_{11}\\
\nonumber &+\star dN_1 \cos{\xi}\sin{\upsilon}+\star dN_1 \cos{\upsilon}\sin{\xi}\\
\partial_{[i}{e_{j]}}^{11}=&\epsilon_{ijk}\partial_k (1-N_1) \cos{\xi}\cos{\upsilon}\equiv \vec{\nabla} \wedge \vec{A} 
\end{align}
The metric is corrected to read:
\begin{align}
\nonumber ds^2=&(N_1N_2N_3)^{\frac{1}{3}}(dx_1^2+\ldots dx_3^2+N_1^{-1}(-dt_4^2+\ldots+dx_{6}^2)\\
&+N_2^{-1}(dx_7^2+dx_8^2)+N_3^{-1}(dx_9^2+dx_{10}^2)\\
\nonumber &+(N_2N_3)^{-1}(dx_{11}+(1-N_1)\cos{\xi}\cos{\upsilon}(dx_1+dx_2+dx_3))^2)
\end{align}
When $\upsilon=\frac{\pi}{2}$ (respectively $\xi=\frac{\pi}{2}$), such that $N_3=1$ ($N_2=1$) the solution reduces to the dyonic membrane. In the limit $\upsilon=\xi=\frac{\pi}{2}$, so that $N_2=N_3=1$, we recover the membrane solution. There are similar limits which give the KK6 solution ($\upsilon=\xi=0$ $N_1=N_2=N_3$) and two different M5 brane limits (1. $\xi=\frac{\pi}{2}, \upsilon=0$ $N_3=1$, $N_1=N_2$, and 2. $\xi=0, \upsilon=\frac{\pi}{2}$ $N_2=1$, $N_1=N_3$. We indicate these limits pictorially in figure \ref{decomposedkk6}. 

There is also a solution which we have not seen before that interpolates between the KK6 brane and an M5 brane, which occurs when either $\xi=0$ or $\upsilon=0$. Let us consider the example where $\upsilon=0$ and hence $N_3=N_1$. The volume element is:
\begin{align}
\nonumber ds^2=&N_1^\frac{2}{3}N_2^{\frac{1}{3}}[(dx_1^2+\ldots dx_3^2)+N_1^{-1}(-dt_4^2+\ldots+dx_{6}^2)\\
&+N_2^{-1}(dx_7^2+dx_8^2)+N_1^{-1}(dx_9^2+dx_{10}^2)\\
\nonumber &+(N_1N_2)^{-1}(dx_{11}-(1-N_1)\cos{\xi}(dx_1+dx_2+dx_3))^2]
\end{align}
The field strength becomes:
\begin{align}
F_{4}=&\star d(N_1)\sin{\xi}
\end{align}

There are another two simple limits of the decomposed KK6 solution that we have not considered hitherto, these are indicated by the diagonals in figure \ref{decomposedkk6}. Upon setting $\upsilon=\xi$, so that $N_2=N_3$ we find that the volume element of the solution becomes:
\begin{align}
\nonumber ds^2=&N_1^\frac{1}{3}(\sin^2{\xi}+N_1\cos^2{\xi})^{\frac{2}{3}}[(dx_1^2+\ldots dx_3^2)+N_1^{-1}(-dt_4^2+\ldots+dx_{6}^2)\\
&+(\sin^2{\xi}+N_1\cos^2{\xi})^{-1}(dx_7^2+dx_8^2+dx_9^2+dx_{10}^2)\\
\nonumber &+(\sin^2{\xi}+N_1\cos^2{\xi})^{-2}(dx_{11}-(1-N_1)\cos^2{\xi}(dx_1+dx_2+dx_3))^2]
\end{align}
The field strength becomes:
\begin{align}
\nonumber F_{4}=&d(N^{-1}_1)\sin^2{\xi}\wedge dt_4\wedge dx_5\wedge dx_6\\
&-\frac{i}{N_2^2}d(N_1)\sin^2{\xi}\cos{\xi} \wedge dx_7\wedge dx_{8}\wedge dx_{11}\\
\nonumber &-\frac{i}{N_3^2}d(N_1)\sin^2{\xi}\cos{\xi} \wedge dx_9\wedge dx_{10}\wedge dx_{11}\\
\nonumber &+\star_2 dN_1 \frac{1}{2}\sin{2\xi}+\star_3 dN_1\frac{1}{2} \sin{2\xi}
\end{align}
The second limit worthwhile considering relates two M5 branes, and appears when we set $\upsilon=\frac{\pi}{2}-\xi$. In this limit, $\sin{\upsilon}=\cos{\xi}$, $\cos{\upsilon}=\sin{\xi}$ so that $(1-N_3)=(1-N_1\sin^2{\xi})$ and, 
\begin{equation}
1-N_3=N_2-N_1
\end{equation}
Our construction prohibits us from taking the limit $N_1\to 1$, corresponding to the vanishing of the M2 charge. However as $N_1$ approaches $1$ the above relation indicates that the charges of the two $S2$ branes  ($(1-N_2)$ and $(1-N_3)$) are approaching opposite values. The parameter $\xi$ interpolates between two M5 branes whose longitudinal directions are $\{t_4,x_5,x_6,x_7,x_8,x_{11}\}$ and  $\{t_4,x_5,x_6,x_9,x_{10},x_{11}\}$ hence its variation in this limit has the effect of rotating a two-dimensional subspace of an M5 brane: $\{x_7,x_8\} \to \{x_9,x_{10}\}$ as $\xi=0 \to \frac{\pi}{2}$. The volume element is:
\begin{align}
\nonumber ds^2=&N_1^\frac{1}{3}(N_1+\frac{1}{4}\sin^2{2\xi}(1-N_1))^{\frac{2}{3}}[(dx_1^2+\ldots dx_3^2)+N_1^{-1}(-dt_4^2+\ldots+dx_{6}^2)\\
&+(N_1+(1-N_1)\sin^2{\xi})^{-1}(dx_7^2+dx_8^2)\\
\nonumber &+(N_1+(1-N_1)\cos^2{\xi})^{-1}(dx_9^2+dx_{10}^2)\\
\nonumber &+(N_1+\frac{1}{4}\sin^2{2\xi}(1-N_1)^{-1}(dx_{11}-\frac{1}{2}(1-N_1)\sin{2\xi}(\sum_{i=1}^3dx_i))^2]
\end{align}
\end{subsection}
\end{section}
\begin{section}{Discussion}
In the decomposition of $E_{11}$ into tensors of $SL(11)$ the gauge fields of the M2 brane and M5 brane appear without discrimination together with an infinite set of mixed symmetry tensors whose direct role in eleven dimensions is unclear\footnote{Although an $E_9$ multiplet \cite{Englert:2007p605} (corresponding to a subset of the $E_{11}$ Young tableaux with a maximum height of nine boxes) has been shown to solve the supergravity equations of motion.}. In this paper we have provided evidence that the mixed symmetry tensors of $E_{11}$ may be interpreted as bound states of M-branes in M-theory in the form of a group element (\ref{interpolatinggroupelement}) encoding these solutions. The group element was paramaterised by two continuous variables: the radial distance $r$ from the origin of the solution in spacetime and an angle variable that interpolated between brane solutions. The angle variable in ranging from $0$ to $\frac{\pi}{2}$ moves along a path connecting the generators of $E_{11}$.

The transversely boosted M2 and M5 branes as well as the dyonic membrane fell into a class of solutions encoded in the group element (\ref{interpolatinggroupelement}). The gauge fields appearing in these solutions mirrored the generators used to move along a path in the adjoint representation of $E_{11}$ under the adjoint action of the algebra. However we also gave examples of the $M2_6$, $M5_3$ and $WM_7$ solutions whose interpolating group element deviated from (\ref{interpolatinggroupelement}). In these examples the orientation of the three component branes was chosen so that the generators associated with their gauge fields formed an abelian sub-algebra. The $M2_6$ and the $M5_3$ were derived using a two-parameter group element whose gauge parameter was modified from the form of (\ref{interpolatinggroupelement}). Interestingly in these examples it was made manifest that given a mixed symmetry Young tableau there is more than one way to decompose it into sub-tableaux before finding a composite solution. For two parameter group elements this point was examined again by reinterpreting the KK6 monopole solution as a bound state of an M5 brane and an S2 brane, although without choosing generators which formed an abelian sub-algebra. Finally a further decomposition of the KK6 monopole into three membrane solutions gave a suggested three parameter group element. In general at level $l$ we expect to use an $l$-parameter\footnote{Where there is one radial parameter, $r$, and $l-1$ interpolating angle variables.} interpolating group element to describe the solution as a superposition of membrane states. Diagrammatically one may imagine the interpolating solutions as an $l-1$-dimensional hypercube, having $l$ vertices corresponding to M2 or S2 branes, the transition along the edges of the hypercube correspond to varying one angle parameter. In this way one may imagine all the states of the adjoint representation as bound states of membranes, or membrane molecules. 

We have concentrated, in this paper, on investigating M-theory bound state solutions about which little is known. A strong test of the form of the group element given in (\ref{interpolatinggroupelement}) is that upon dimensional reduction is gives rise to other well-known bound state solutions. In particular in a forthcoming work \cite{Cook:2009p2739} we intend to reproduce the bound states of string theory by a similar analysis. There are some interesting questions associated with this reduction in particular the question of whether we can use the group element formulation to produce localised bound state brane solutions. Specifically bound states which include D6 branes are known to give localised brane solutions \cite{Cherkis:2002p1984} as opposed to the smeared ones we have considered in the present work, since they arise as a dimensional reduction of  a pure gravity solution (as a reduction of the KK6 monopole). In this paper and elsewhere \cite{Englert:2007p605} an infinite tower of gravitational solutions have been derived and it is interesting to wonder whether other localised brane solutions may be derived by dimensionally reducing the bound states of this tower of solutions.

Amongst the examples considered were the transversely boosted M2 and M5 branes the interpolating solution parameterised the path from ${K^a}_b$ to $R^{acd}$. In the example shown in section 4.2 the action $[{K^8}_{11}, R^{91011}]=R^{8910}$ was encoded in the group element and a rather complicated metric was derived which included the angle parameter, $\xi$. In \cite{Russo:1996p693} this angle parameter was interpreted as the boost parameter of the Lorentz transformation associated to the rotation generator ${K^8}_{11}$ in $SO(1,10)$ and by observing that the resulting field strength should simplify to an M2 brane field strength written in Lorentz boosted coordinates it was possible to read off the Lorentz transformations on the coordinates. In an identical approach one would expect to be able to interpret the action of the $R^{abc}$ generator on the coordinates by considering the field strength of the dyonic membrane. In this case the problem is not so straightforward since the transformation should map an involved four-form field strength to a simple seven-form field strength of the fivebrane. One can postulate that the map acts on the coordinate three forms as:
\begin{align}
\nonumber dx_6\wedge dx_7\wedge dx_8 &\to \frac{1}{\sin{\xi}}\sqrt{\frac{N_1}{N_2}}\wedge d\Omega\\
dx_9\wedge dx_{10}\wedge dx_{11} &\to -i\frac{1}{\tan{\xi}}\sqrt{\frac{N_2}{N_1}}\wedge d\Omega
\end{align}
So that the four-form field strength is transformed into a seven-form field strength:
\begin{equation}
{\cal{F}}_{4(M2\oplus M5)}\to d\frac{1}{\sqrt{N_1N_2}}\wedge d\Omega
\end{equation}
Where $d\Omega=-d\bar{t}_6\wedge d\bar{x}_7\wedge \ldots \wedge d\bar{x}_{11}$ is a six-form and $\bar{x}_i$ are the transformed coordinates whose transformation bring the volume element to:
\begin{align}
d\bar{s}_{M2\oplus S2}^2&=({N_1N_2})^{\frac{1}{3}}(dx_1^2+\ldots dx_5^2+{(\sqrt{N_1N_2})}^{-1}(-d\bar{t}_6^2+d\bar{x}_7^2+\ldots + d\bar{x}_{11}^2))
\end{align}
However finding the precise nature of the coordinate transformations as was possible with the Lorentz transformations, corresponding to the action of $R^{abc}$ remains an open problem.

By relating the high level roots of $E_{11}$ to composite bound state solutions we hope to gain an understanding of the field equations to be satisfied by exotic brane solutions. For the low level single membrane field strength the equation of motion does not depend upon the Chern-Simons term of the bosonic part of the supergravity action. The dyonic fivebrane equations of motion in D=8 are dependent upon the dimensionally reduced Chern-Simons term; the curved space Laplace equations in D=8 are sourced. In eleven dimensions the fivebrane equations of motion are not generically the same form as the membrane equations of motion, there is a contribution to the equations of motion from the Chern-Simons term. Understanding the transformation of $A_{[3]}$ and ${\cal F}_{[4]}$ under the action of $R^{abc}$ ought to determine the transformed Lagrangian including the transformation of the Chern-Simons term and the equations of motion. The corresponding Lagrangian will deviate from a sum of lagrangians associated to the constituent branes by additional terms coming from both the kinetic terms and the Chern-Simons terms and the difference corresponds to the binding energy of the system. 

The dyonic membrane is a solution to the equations of motion for arbitrary angle parameter and not just $0$ or $\frac{\pi}{2}$. That it falls into the class of bound state solutions encoded in the group element of \ref{interpolatinggroupelement} suggests that one must consider the full and continuous $E_{11}$ symmetry as a symmetry of M-theory and not only the discrete subgroup generated by U-duality transformations.

\end{section}
\section*{Acknowledgements}
It is a pleasure to thank Laurent Houart, Neil Lambert, Ergin Sezgin, Per Sundell and Peter West for various discussions relating to aspects of this work. In particular I am grateful to Axel Kleinschmidt for his interesting and helpful comments which improved the original version of this paper. In addition I would like to thank the Scuola Normale Superiore di Pisa for its hospitality during the completion of this work.

\bibliographystyle{utphys}
\bibliography{KKbranesref}

\end{document}